\documentclass[12pt,preprint]{aastex}
\slugcomment{Accepted  for publication
in {\it The Astrophysical Journal}}
\def\lax {\ifmmode{_<\atop^{\sim}}\else{${_<\atop^{\sim}}$}\fi}
\def\gax {\ifmmode{_>\atop^{\sim}}\else{${_>\atop^{\sim}}$}\fi}
\def\gtorder{\mathrel{\raise.3ex\hbox{$>$}\mkern-14mu
             \lower0.6ex\hbox{$\sim$}}}

\newcommand{\kal}{K$_\alpha$}
\begin{document}

\title{On the Non-relativistic Origin of Red-skewed Iron Lines in CV,
Neutron Star and Black Hole Sources}

\author{Lev Titarchuk\altaffilmark{1}, Philippe Laurent\altaffilmark{2} \&
Nikolai Shaposhnikov\altaffilmark{3,4}
}

\altaffiltext{1}{George Mason University, Fairfax, VA 22030;  US Naval Research
Laboratory, Code 7655, Washington, DC 20375-5352; 
Goddard Space Flight Center, NASA,  code 663, Greenbelt
MD 20771, USA and Physics Department, University of Ferrara, Via Saragat, 1
44100 Ferrara, Italy; lev.titarchuk@nrl.navy.mil}
\altaffiltext{2}{CEA/DSM/DAPNIA/SAp, CEA Saclay, 91191 Gif sur Yvette, France;plaurent@cea.fr; Laboratoire APC, 10 rue Alice Domont et Leonie Duquet, 75205 Paris Cedex 13, France}
\altaffiltext{3}{University of Maryland, Astronomy Department, College Park, MD 20742}
\altaffiltext{4}{CRESST/GSFC/NASA, code 662, Greenbelt MD 20771}

\begin{abstract}
We perform the analysis of the iron K$_{\alpha}$ lines  detected in three sources
representing of three types of accreting compact sources: 
cataclysmic variable (CV) GK Per, neutron star (NS) Serpens X-1 and black hole (BH) GX 339-4. 
We find,
using data from Epic-PN Camera on-board XMM-{\it Newton} observatory,
that  the iron  K$_{\alpha}$ emission line in GK Per has a noticeable
red-skewed profile.
We compare the GK Per asymmetric line with  the red-skewed lines observed by
XMM-{\it Newton} in  Serpens X-1 and GX 339-4.
The observation of the K$_\alpha$ emission with red-skewed features
in CV GK Per 
cannot be related to the redshift effects of General Relativity (GR).
Therefore, if the mechanism of the K$_{\alpha}-$line formation  is the same in CVs, NSs and BHs  then
it is evident that the GR effects would be ruled out  as a cause of red skewness of K$_\alpha$  line.
The line reprocessing in an outflowing wind has been recently suggested  an alternative model for a broad 
red-shifted iron line formation. In the framework of the outflow scenario
the red-skewed iron line is formed in the strong extended 
 wind  due to its illumination 
by  the radiation  emanating from the innermost part of the accreting material.
The outflow is a common phenomenon for CVs, NSs and BHs.  
In this Paper we demonstrate that the asymmetric
shapes of the lines  detected from these CV, NS and BH sources  are well described with
the wind (outflow) model. 
Furthermore we find by analyzing {\it RXTE} observations
that when the strong  red-skewed iron line is observed in GX 339-4 
the high frequency variability is strongly suppressed.  
While this fact is hard to reconcile with the
relativistic models, it is consistent with the outflowing gas
 washing out high frequency modulations of the radiation presumably originated  in the innermost part of the source.

\end{abstract}

\keywords{line: profiles ---stars: white dwarfs, neutron ---X-ray :binaries --- X-rays: individual (GK Per), individual (Serpens X-1), individual (GX 339-4) ---radiation mechanisms---physical data and processes}

\section{Introduction}
The \kal\ iron line is emerging as one of the most important probes
of physical processes near accreting compact objects, which include black
holes (BH), neutron stars (NS) and white dwarfs (WD). Strong red-skewed iron
lines have been reported in multiple observations of BHs \citep{mill07}
and CVs \citep{hm04}, hereafter HM04.
With a discovery 
of a red-skewed iron line in the NS source Serpens X-1 by \citet{bs07}, hereafter BhS07,
the asymmetric lines have now been observed in all types of accreting compact objects.
In this Paper we analyze the line profiles in CV (WD) GK Per, NS Serpens X-1 and BH GX 339-4 
as representatives of each compact object sub-class. The reason for such a study is simple:
if the Physics leading to the asymmetry of the iron line profile in all types of compact objects is the same, it will 
allow us to put additional constrains on the iron line formation models.

HM04 presented the compilation of Fe K$_\alpha-$line spectra of  5 magnetic CVs
using the Chandra  High-Energy Transmission Grating (HETG).
The GK Per fluorescent line had the highest signal-to-noise ratio. HM04 demonstrated that the \kal\ iron line  in GK Per showed a Doppler-shifted red
wing extending to 6.33 keV which they attribute to pre-shock material  falling with at near WD escape velocity.  They also detected a fainter, 170-eV shoulder which, they suggested, may be due to Compton downscattering.

BhS07 reported on an analysis of XMM-Newton data from  NS
low-mass X-ray binary (LMXB) Serpens
X-1 (Ser X-1). Their spectral analysis of EPIC PN data indicates that the
previously known broad iron \kal\ emission
line from this source has a  skewed structure with a
moderately extended red wing.
The BhS07 finding suggests that the broad lines seen in other NS
 LMXBs would likely have  the skewed shape as well, the origin
of which is still under debate. In fact, BhS07 showed that the
quality of fits of observed red-skewed lines in NSs are the same for
both the ``diskline'' model [see \cite{fab89}] related to
Schwarzschild potential and for the  ``Laor'' model [see \cite{laor}]
related to a maximally spinning BH. It implies that the
red-skewed line is not a particular signature of a maximally
spinning black hole given that this shape is also common to the
lines from NS sources. Subsequent {\it Suzaku} observation of three
NS LMXBs (Ser X-1, 4U 1820-30 and GX 349+2) analyzed by \cite{c07}
have shown broad, asymmetric, iron
emission lines in all three
objects. These iron K$_{\alpha}$ lines can be also well-fitted by the model for
lines from a relativistic accretion disk (diskline), and by the
`Laor' model.
\citet{mill07} reviewed observations of red-shifted line from
BH sources and their interpretation in terms of the relativistic 
models.

The iron
fluorescence related to  the reflection of hard (power law) spectrum from the cold accretion disk 
have been the most widely studied model for asymmetric emission line formation [see \cite{rn03} and
\cite{mill07}]. However, all theoretical alternatives have to be  explored and properly 
dismissed before indisputably claiming of such an
important physical effect as General Relativity (GR) as an origin 
of this redskewed line phenomenon. As we show in this Paper, the GR interpretation of the red-skewed lines
in NSs and BHs is not an unique model which successfully  describes the observed line profiles. 
An alternative  viable explanation of the line properties can be achieved in terms of 
downscattering of line photons in the wind outflows which are commonly produced in these 
systems.

A model for  red-skewed  emission lines by repeated electron
scattering in a diverging outflow or wind was considered  by
\cite{t03},  \cite{lat04}, and \cite{lt07}, hereafter LT04, LT07
respectively. Let us remind the reader  the geometry of the wind model.
 It is assumed the wide-open wind is
launched  at some disk radius $R_{W}$ where presumably the local
radiation force $\sigma L(R_{W})/c$  exceeds the local
disk gravity $m_pHGM/R_{W}^3$ [see e.g. \cite{t07}].  Here $L(R_{W})$ is the disk local
luminosity at $R_{W}$, $\sigma$ is an effective plasma cross-section
which is, in principle, greater than the Thomson electron
cross-section $\sigma_{\rm T}$; $H$ is the half-height of the disk,
and $m_p$ is the proton mass.  Thus this wind should be illuminated
by the emission of X-rays formed in the innermost part of the
source. The K$_{\alpha}-$line is generated in a  narrow wind
shell and the line profile is formed in the partly ionized wind 
when K$_{\alpha}-$line photons are scattered  off the
diverging flow (wind) electrons. Electron scattering of the Fe K$_\alpha$ photons within the highly ionized expanding
flow leads to a decrease in photon energy (redshift), which is of the first order in
$\beta_{W}=V_{W}/c$, when $V_{W}$ is the outflow velocity with $V_{W} \ll c$.
This photon redshift is an intrinsic property of any
outflow for which divergence is positive.

LT07 fitted the Monte-Carlo (MC) model line profiles to the observations
using just three free parameters: $\beta_{W}$,  the optical depth of the wind $\tau_W$ and the wind plasma temperature $T_{W}$.
LT07 showed examples of the outflow model
 fitted to the XMM observation
of MCG 6-30-15 \citep{W01} and to the ASCA
observations of GRO J1655-40 \citep{m04a}.
The red-skewed part of the spectrum is formed by photons undergoing
multiple scatterings while the
primary peak is formed by photons escaping directly to the observer.




There are two completely independent lines of reasoning motivating a
consistent study of the asymmetric Fe lines in WDs, NSs and BHs.
The first comes from the fact that certain theoretical alternatives to relativistic disk
interpretations in the BH and NS sources, namely those alternatives
based on wind outflow geometries, are not restricted to those
classes of accreting compact objects but could readily include
accreting cataclysmic variables (CV), meaning an appropriate subset
of WDs.
However, there
is a second point which is more general.  Just as the NS case cannot
be modeled appropriately with a near-maximal Kerr metric because the
NS precludes such a high angular momentum, a WD case would have to
be modeled with a disk whose gravitational potential was effectively
Newtonian, far from relativistic regions of the Schwarzschild
metric. Thus, within the context of the more popular disk models and
without any reference to any alternative models such as outflow, it
would force a re-examination of some of the basic assumptions.

We analyze the K$_{\alpha}$ lines
observed in the GK Per, Ser X-1 and GX 339-4 with XMM-{\it Newton}.
For GX 339-4 we also analyze simultaneous {\it RXTE} data.
We model the line profiles both in terms of relativistic models
and in terms of wind/outflow paradigm. Additionally, we consider GX 339-4 XMM-{\it Newton}/{\it RXTE}
observations in terms of broader context of line behavior with respect 
to fast source variability which provides  additional information on
the system geometry and allows to constrain line formation models.

We present observations and the data analysis in \S 2. In \S 3 we discuss our results
and their implications to theoretical models. Conclusions follow in \S 4.

\section{Observations and Data Analysis}

The primary instrument for our study of the iron line profiles is the XMM-{\it Newton}/Epic-PN 
camera. We also use {\it RXTE} observations as  supporting spectra and timing data.
The corresponding data sets were downloaded from HEASARC\footnote{http://heasarc.gsfc.nasa.gov/} and analyzed following the ``SAS User's Guide'' using {\tt especget} SAS tool.   
For GK Per we use XMM observation on March 9 2002 (ID 0154550101/revolution 412).
The Epic cameras were operating in the full imaging mode during this
observation. We extracted the source spectrum from a circular region
centered on the GK Per position with a radius of 35 arcsec and we 
extracted a background spectrum from a nearby source free region.
In addition, we present the {\it RXTE} spectrum of GK Per from the 
observation on March 8, 2002 (Obs ID 70401-01-02-00) which was the closest to that by XMM.

We use  the XMM-{\it Newton} observations
of NS Ser X-1  (XMM rev. 785), and BH GX 339-4 
(see  \S 2.3
for details of  the XMM/{\it RXTE} observations). 
XMM/Epic-PN  data for Ser X-1 and GX 339-4 were collected in the timing
mode, for which spacial information is available for one dimension (RAWX) only.
We extracted source and background spectra from stripes according to selection
criteria 10$<$RAWX$<$60 and 2$<$RAWX$<$9 correspondingly. 
For the GX 339-4 
observations we also used simultaneous {\it RXTE} data.
We extracted PCA and HEXTE energy spectrum from
standard {\it RXTE} data modes. For spectral analysis  we use  the energy ranges
1.0-10.0, 9.0-30.0 and 20.0-200.0 keV for {\it XMM}/Epic-PN, {\it RXTE}/PCA and  {\it RXTE}/HEXTE
spectra correspondingly. {\it RXTE} response matrices were calculated using
XTE FTOOLS package and  ``RXTE Cook Book'' recipes. 
For timing analysis we  calculated Fourier 
power density spectrum (PDS) from the high timing resolution {\it RXTE} data
modes in the frequency range 0.05 to 256 Hz. 
 
Spectral analysis was performed using XSPEC astrophysical spectral modeling package.
The models applied to the data and the results of spectral fits are shown
in Tables \ref{gkper_tab}-\ref{gx339_sept292002_tab}.  Spectral shapes for 
observations of GK Per, Ser X-1 and GX 339-4 (Observations 1 and 2) indicate 
the presence of a strong iron emission line.  We apply three models for
 the iron line: DISKLINE, LAOR and WINDLINE. Note DISKLINE and LAOR are
 standard XSPEC  models. The DISKLINE model describes the line emission produced by a relativistic 
accretion disk \citep{fab89}. The LAOR component presents the line from an accretion disk around a black hole including GR effect calculated by \citet{laor}. The WINDLINE model is a Monte-Carlo
 code simulating the downscattering of the iron line photons in a diverging wind from a compact object
 as per \citet{lt07}. The code was ported into XSPEC as a local model.

\subsection{GK Per}

Cataclysmic variables (CVs) are interacting binaries in which the
accreting object (the primary) is a white dwarf [see \cite{w95} for
a review]. X-ray emission in CVs is most likely associated with the
accretion process, which is capable to heat  accreted
material at a shock up to high temperatures ($kT \sim$5-20 keV). Thus one
can fit the continuum spectrum of GK Per emission by the model in
which the soft photons originated in the accretion disk (or at the WD
surface) are upscattered (Comptonized) off hot electrons of the
plasma cloud (shock region). 

\begin{figure}[ptbptbptb]
\includegraphics[scale=0.7, angle=-90]{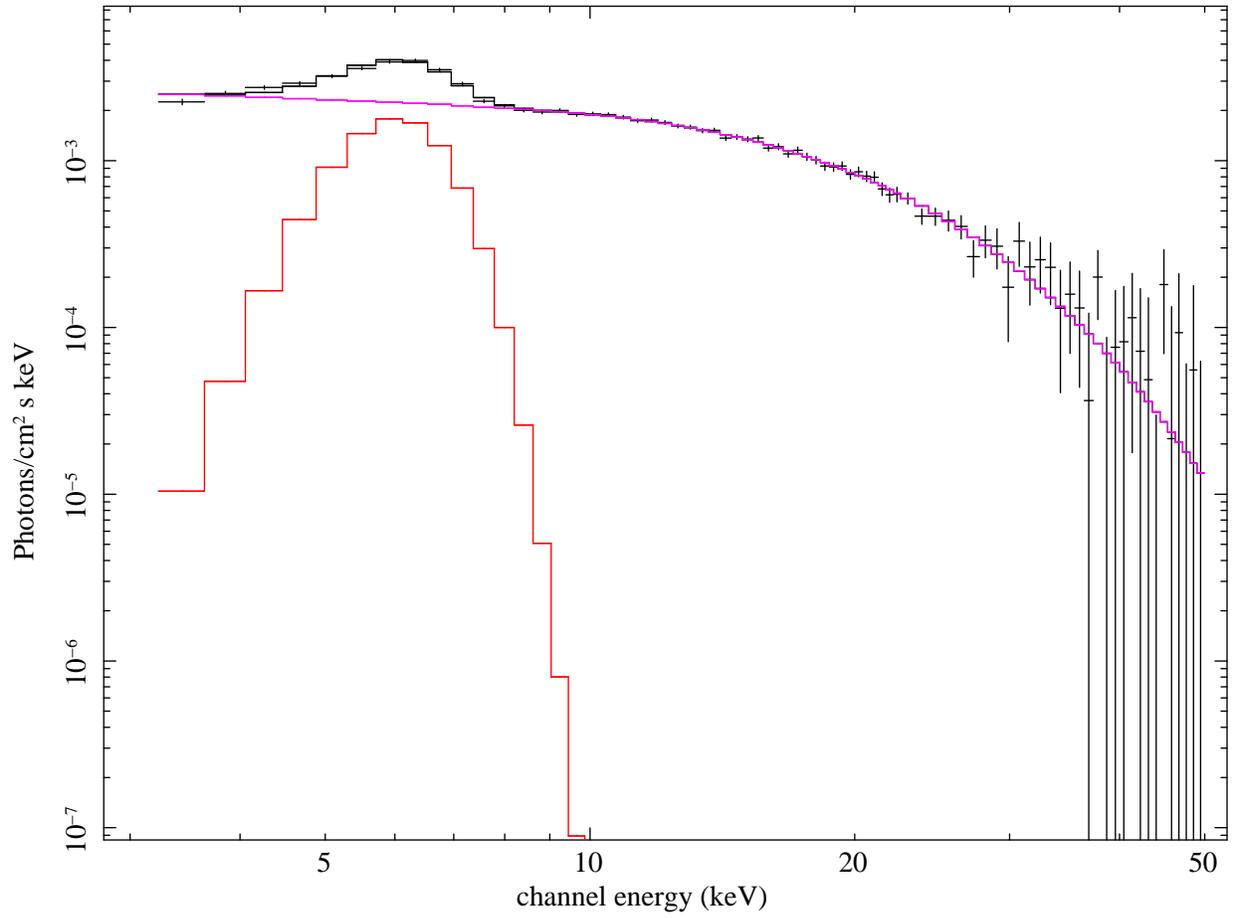}
\caption{The {\it RXTE}  observation of  GK Per  
The data (error bars) vs model (solid line).  Here we have used the best-fit  model  WABS$\ast$(COMPTT+GAUSSIAN).}

\label{GKPER_fit_cont}
\end{figure}

Given that,  we use the Comptonization model by
\cite{t94} and \cite{ht95} (see COMPTT model in XSPEC)  to infer the
electron  temperature $T_e$ and the optical depth $\tau_e$ of the
plasma cloud where  the emergent spectrum of GK Per is produced. 
Unfortunately, there is no simultaneous observations of GK Per by  XMM and {\it RXTE} therefore
we choose the nearest {\it RXTE} observations to estimate the model for continuum emission which we 
then can use as a starting model to fit the XMM data.  In Figure \ref{GKPER_fit_cont} we present the {\it RXTE}/PCA spectrum of GK Per 
fitted by the model  WABS$\ast$(COMPTT +GAUSSIAN). The best parameters of the model are hydrogen  column  
$n_H\sim1.5\times10^{22}$ cm$^{-2}$ (frozen),
color temperature of the seed photon of COMPTT $T_0=0.5$ keV (frozen),
plasma temperature of the Compton cloud (CC) $kT_e=5.1\pm 0.2$ keV, CC optical depth
$\tau_e=25\pm 5$, the energy of the Gaussian line $E_L=6.1\pm 0.3$ keV, the line width 
$\sigma=0.8\pm0.1$ keV. The model is acceptable given that $\chi^2= 73.6/77=1.04$.
The above COMPTT best-fit parameters correspond 
to  the spherical geometry of Compton cloud. 

HM04 reported the results of  their analysis  of {\it Chandra} HETG data  from GK Per. Specifically they pointed out that   Fe K$_{\alpha}$  fluorescent line  exhibits an extended red wing.
The Chandra spectrum of GK Per (see Fig. 1 in HM04) may also indicate  
the  presence of 6.7 keV  and weak  6.96 keV lines, that is  presumably the Fe$_{XXV}$  (He-like) and 
 Fe$_{XXVI}$ (H-like) lines respectively. 
 In our analysis we confirm that the  Fe K${_\alpha}$ line in GK Per is asymmetric. 
However, XMM data do not show any evidence for  the  6.7  and 6.96 keV narrow lines (see Table \ref{gkper_tab}).

\begin{figure}[ptbptbptb]
\includegraphics[scale=0.7, angle=-90]{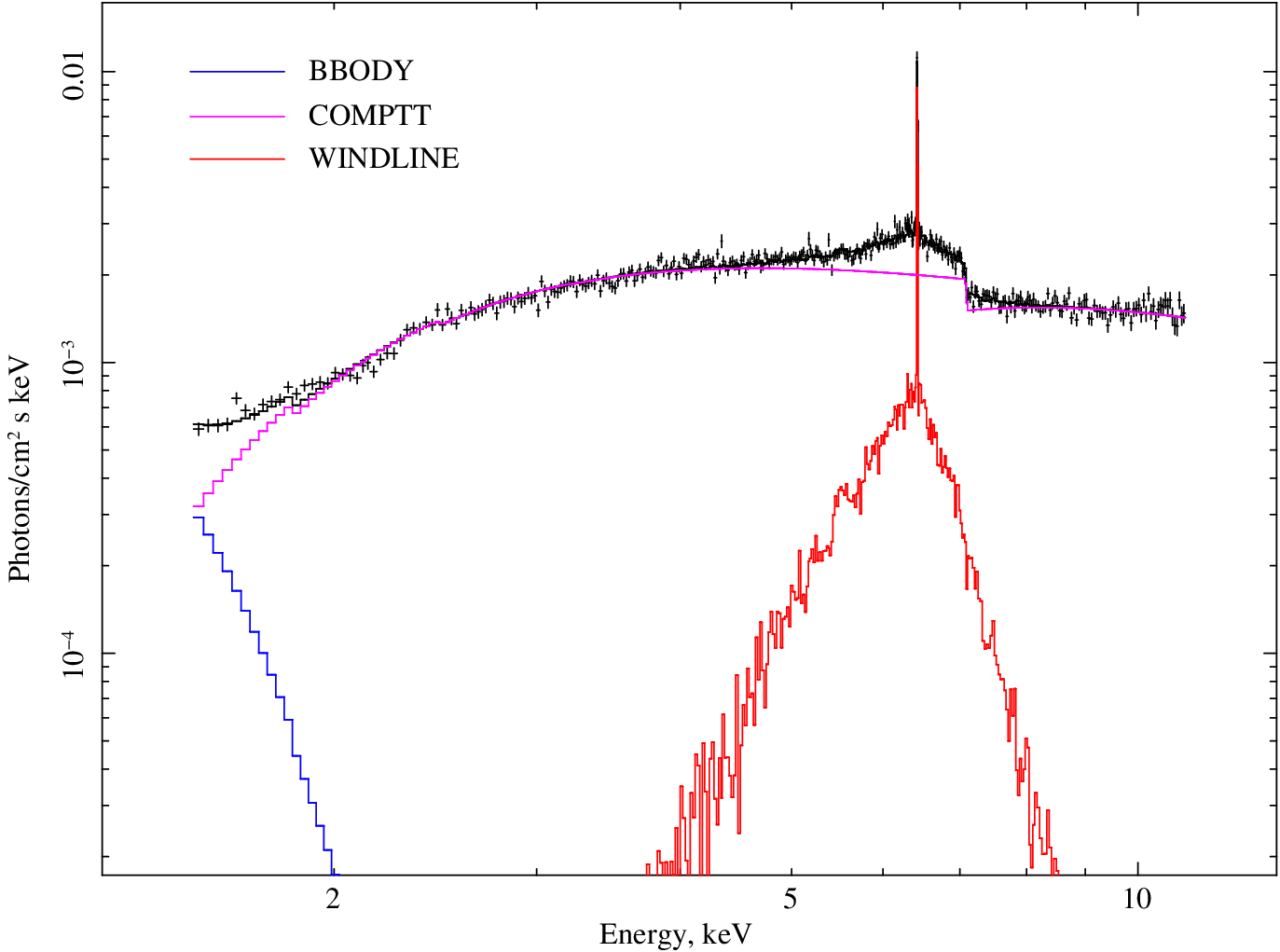}

\includegraphics[scale=0.705, angle=-90]{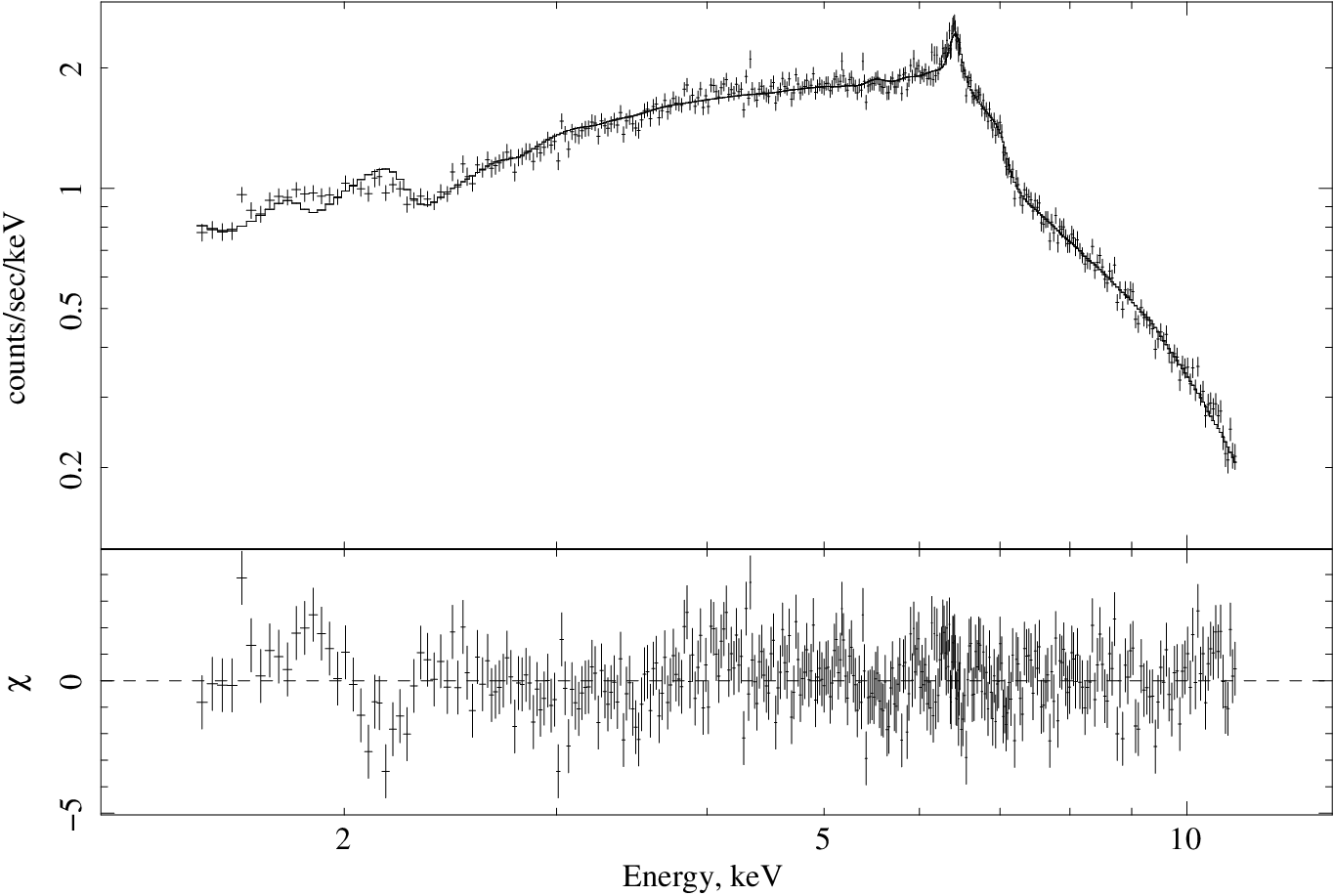}
\caption{The XMM-{\it Newton} observation of  GK Per  (rev. 412) 
The data (error bars), model (solid line).  Here we have used the best-fit  model wabs$\ast$(COMPTT+BBODY+WINDLINE)$\ast$EDGE (see Table 1).
}
\label{GKPER_fit}
\end{figure}

We first tried to fit the XMM-{\it Newton} spectrum from GK Per with the model
WABS$\ast$\-(BBODY + COMPTT)$\ast$EDGE. We find that  the quality of the fit using
 this model is unacceptable, $\chi^2/dof= 2.4$ and also we find a strong red-skewed residual in the 
energy range between 5 and  7 keV. Addition of the GAUSSIAN with the energy $\sim$ 6.4 keV 
leads to the improvement in the fit quality to $\chi^2_{red}=1.5$, which is still unsatisfactory.
This indicates that the line profile is not symmetric. 
We then replaced GAUSSIAN with a physically motivated model to fit the asymmetric iron line,
i.e. DISKLINE, LAOR or WINDLINE model.  We were able to obtain acceptable fit quality for
all models. However, the best fit statistic  of  $\chi^2/dof=1.14$ is given by the WINDLINE model.
 The quality of the model fit to the data for the WINDLINE model is also shown in Figure \ref{GKPER_fit}
 (lower panel) where  we present the count spectrum  (error bars) vs model (solid line) along with the 
residual in units of 
$\chi-$deviation of the model  from the data,  i.e.  $\chi=\rm{(data-model)}/\sigma$.  
 In the upper panel of Figure \ref{GKPER_fit} we show the resulting photon spectrum along with 
its continuum and line components.  The red wing of the line extended from 6.4 keV down to 5.8 keV 
(half width of half maximum HWHM$\sim 0.6$ keV)  is broader than the blue wing that HWHM of 0.3 keV.
The same model fit with the LAOR model is shown on Figure \ref{GKPER_fit_laor}.
In Table   \ref{gkper_tab}  we present the best-fit parameters 
for the continuum and line components for all three model fits.

Modeling the XMM continuum spectrum of GK Per with COMPTT yields the electron temperature
 $kT_e=5.3$ keV and Thomson optical depth $\tau_e= 20.1$, which are consistent with 
{\it RXTE}  spectral parameters. The reported value of  the optical depth is for the spherical geometry. 
The best-fit absorption optical depth  of the multiplicative K-edge component  $\tau_{max}=0.23$.
The  K-edge energy range starts at 7.1 keV. 


We apply DISKLINE and LAOR models to fit the line profile only formally to
 demonstrate that the XMM data from GK Per can be successfully fitted with these relativistic line models. 
The XSPEC DISKLINE and LAOR models 
were designed to fit  to  the  observable 
red-skewed line profile which is assumed to be a result of 
the General Relativity (GR) effects [see e.g. \cite {TAN}]. 
However,  the GR  effects are negligible  in the WD case and thus the application of the GR models to fit 
the red-skewed iron lines of GK Per {\it  cannot provide physically meaningful  results}. 
The best-fit values of the disk inner radius are $R_{in}=12.5, ~22R_{G}$  for LAOR and DISKLINE models respectively. These values are 3 orders of magnitude less than the WD radius that is about $10^{9}$ cm.
In fact, we present   
the fitting results for WD source  using the GR models  to demonstrate that 
{\it a good quality fit of the line profile by the  relativistic models can not be 
considered as a reliable observational evidence of the  GR effects.}
This fact has an important consequence for the results of application of the relativistic
models to NS and BH spectra. Namely, a success of these models solely based on the GR effects
to fits the iron line profiles in NS and BH systems cannot serve as a solid grounds for
claims that these effects are  at work in these systems.

On the other hand  the WINDLINE model is  a physically justified model for the line production in GK Per. 
 This  model  includes both scattered and 
direct components of the line. The free parameters of the wind model are the dimensionless wind 
velocity $\beta_{W}=V_{W}/c$, Thomson optical depth of the wind $\tau_{W}$,  plasma temperature  of the wind $T_{W}$ and  the line energy $E_{L}$.
The fit of  the model  [WABS$\ast$(BB + COMPTT \ +WINDLINE)$\ast$ EDGE] to the spectral data of GK Per  observed by Epic-PN XMM  (see Fig. \ref{GKPER_fit} and Table \ref{gkper_tab})  provides  the 
best-fit   values of the wind model components, specifically   $\beta_{W}=(0.71\pm0.06)\times10^{-2}$, $\tau_{W}=1.46\pm0.06$ and 
$kT_{W}=0.97\pm0.06$ keV. Apart from the unphysical outcome 
 produced by DISKLINE and LAOR models,
the best-fit parameters of the WINDLINE models represent a conceivable situation
for an accreting WD system.    

\subsection{Ser X-1}

To resolve   the iron line profile  in EPIC-PN data for Ser X-1 (XMM rev 785) we follow the same procedure that we apply for the analysis of the GK Per data.
These data were analyzed by BhS07 who
first reported the red-skewed line in a NS source. 
They interpreted  the asymmetric shape of the line in terms of
GR line models (LAOR and DISKLINE). Consequently
they  suggested this broad asymmetric line in NS  likely originated
from the innermost part of the disk that was similar  to the
interpretation of the red-skewed lines observed in BH sources.   

We fit  three models to the
data for   which we use LAOR, DISKLINE and WINDLINE components to describe the iron line profile.
 The best-fit values of parameters and goodness of the fit for each model are
shown in Table \ref{serpx1_tab}. 
The fit quality are not significantly different for all models. 
In Figure \ref{ser_fit}  we  present
the iron line profile which we find  in Ser X-1 with the WINDLINE model.
 The upper panel shows the data (error bars) vs model (solid line) 
where we also show each spectral components separately.
The lower panel displays the count spectrum along the residual of the 
data vs model in units of  $\chi=\rm{(data-model)}/\sigma$.


According to our results  the wind velocitiy and
wind optical depth  in Ser X-1   are similar to that of GK Per, namely 
 $\beta_{W}\sim 0.015$ and $\tau_{W}\sim 1.6$.  However, we find that the K$_\alpha-$line energy at 
$E_{L}=6.67$ keV instead of 6.4 keV in the case of GK Per.
The best-fit values of the COMPTT component of the model,  presented in Table \ref{serpx1_tab}, 
the seed photon color temperature $kT_0=0.5$ keV, 
  $kT_e=2.9$ keV  and $\tau_e=13.4$ correspond to  spherical geometry of Compton cloud.

The disk inner radius $R_{in}$ is the most important parameter of the relativistic models.
In our spectral modeling with the LAOR component the best fit value of $R_{in}$ is $1.23$ in units 
of gravitational radius  ($R_g=GM/c^2$), which is the lowest limit allowed by the model. 
BhS07 obtained  that $R_{in}$ varied  between 4.14 and  25 $R_g$ for the LAOR model.
Note that the DISKLINE model  related to the  Schwarzschild background  
is more appropriate for NS  than LAOR model where the Kerr geometry 
is used to describe the curved space around a maximally  rotating central 
object.  For the DISKLINE model we obtained the disk inner radius $R_{in}\sim 6 R_{g}$ which is also the lowest limit value for this parameter set by the model.

\subsection{GX 339-4}

The red-skewed line has been  observed  in the recurrent
Galactic BH transient  GX 339-4  on multiple occasions at different fluxes, in
different states, and with both {\it Chandra} and XMM-{\it Newton}
[see \cite{mill07}, hereafter M07,  for details of these  observations and for the
GR interpretation of the line origin].
In particular, \cite{m04b}, hereafter M04b,  presented  the combined analysis of XMM (rev. 514) 
and {\it RXTE} observations
 during a bright phase of its 2002-2003 outburst.  They reported an extremely skewed  
Fe K$_\alpha$ emission line  in these data. 

For our analysis  we select three XMM-{\it Newton} observations made
 on August 24, 2002, September 29, 2002 and March 8, 2003 which were accompanied by
simultaneous {\it RXTE}/PCA pointing observations (see Table \ref{gx339_tab} for XMM/{\it RXTE} data). 
We further refer to these data sets as Observation 1, 2 and 3 respectively.
Joint XMM-{\it Newton}/Epic-PN and {\it RXTE}/PCA/HE\-XTE spectrum covers  the energy band  from 1 to 200 keV.
{\it RXTE}/PCA high time resolution data allow timing analysis at time scales down to 
milliseconds.

Our primary goal is to study the properties of the iron \kal line.
In order to assess the general behavior of the line we first fit all three
observations using the model in which the line shape is  described by  a simple Gaussian.
The centroid energy of the Gaussian was fixed at 6.4 keV while the width 
was allowed to change. The  fit quality $\chi_{red}^2$ along with the essential parameters of
the model are presented in Table \ref{gx339_tab}. Unfolded energy spectra for
each observation are presented  in Figure \ref{gx339_ga_eufs} (left panels)
accompanied by  the {\it RXTE}/PCA power density spectra (right panels). 
The  Observation 3 spectrum   is  harder than the spectrum 
 for  Observation 2:  the BMC model indices $\Gamma\sim 2.17$ and  
$2.54$ for the former and latter cases respectively. 
According the Homan-Belloni's BH classification [see \citet{hb05}]  during the Observation 3 the source  is in the hard intermediate state (HIMS) while during  Observation 1 ($\Gamma\sim 2.37$) and  Observation  2 the source   is in the soft intermediate state (SIMS) and 
high soft state (HSS) respectively.  

The gaussian fits to the iron line show the strongest line ($EW=684$ keV) during 
Observation 2 while the line is weakest during  Observation 3. 
Note for  Observation 3  the fit by Gaussian  is good giving $\chi^2_{red}=1.02$.
As for Observation 2  the quality of 
the  fit  with a gaussian ($\chi^2_{red}=1.27$) is  worse than that  for Observation 3.  Observation 1 presents an intermediate case between
Observations 1 and 3. This qualitative analysis indicates that during  Observation 2
the line profile is not symmetric and use of more sophisticated physical models 
is required to adequately describe the line shape. For  Observation 3 the gaussian 
fit adequately models the spectrum. Based on the statistical significance
we can conclude that during Observation 3 the line is weak and its profile is symmetric.   
Also given that statistical quality of the data for both Observations 2 and 3 are comparable
we can suggest   
that 
the line skewness  during Observation 3,
is much weaker than that for  Observation 2.  Thus we choose Observation 2 data 
for application of physical models in order to investigate the origin of the line skewness which can be 
either  due  to scattering in a  wind or due to relativistic red-shift effects.


Similarly to the analysis of GK Per and Ser X-1, we apply three different models (LAOR, DISKLINE and WINDLINE) to model  the redskewed iron line seen in  Observation 2  spectrum.
Table \ref{gx339_sept292002_tab} summarizes the resulting  best-values of model parameters. 
In Figure \ref{gx339_514_fit} (see lower panel) we present  the model fit with WINDLINE model for 
which $\chi^2_{red}=\chi^2/(dof)$ has the lowest value of 1.24.

We confirm the presence of the very strong red-skewed 
line extended from about  4 keV to  8 keV  in the data for  Observation 2 
by M04b (see the red-skewed line profile  in upper panel of Fig. 
\ref{gx339_514_fit}).   
However, we  do not confirm  expectations by \cite{m04c}, hereafter M04c, that  the wind velocity of 
$V_{W}\sim 0.3 c$ is required  to fit the strong red-skewed line detected in GX 339-4.
In fact, our  best-fit value of $\beta_{W}=V_{W}/c$ is  0.025. However our results agree with the M04c 
prediction of high optical depth in the wind. We find that the  best-fit value of $\tau_{W}$  
is about 4.9 (see Table \ref{gx339_sept292002_tab}). 
Modeling the spectrum of  Observation 2 using  the LAOR and DISKLINE components
 produces 
values for the inner disk radius of
1.23 $R_{g}$ and 6.0 $R_{g}$ for the LAOR and DISKLINE models respectively. 
As in the case of Serp X-1, these values  are, in fact, the
lowest allowed values  for this parameter in these models. 

Note these low values of $R_{in}$ imply that the observer
should directly see the innermost region of the source during this particular observation.
Thus one can expect to observe high frequency variability from 
the innermost region on accretion flow. However, contrary to this expectation,
the {\it RXTE}/PCA Fourier power spectra show (see right panels on Figure \ref{gx339_ga_eufs}) 
that during Observation 2 the high frequency variability is not present at all in the power density spectrum (PDS), see details in the discussion section.
As we have already pointed out   the line during Observation 3 is weaker 
than that  
during Observations 1 and 2 and the line fit  by Gaussian is  good which implies the symmetry of the line. 



M04b interpreted  the spectral data of  XMM rev. 514 (our Obsercvation 2) in terms of the LAOR   model. 
They  also claimed that  the ionized disk reflection component should be included in the continuum  
model in order to fit the data. In our analysis of the same XMM-{\it Newton}/{\it RXTE} data the reflection 
component is not required.
This difference between our continuum spectrum and that of  M04b  can be explained by  
different components used to represent the non-thermal part of the spectrum. 
We apply the BMC model  while  M04b uses the sum 
of thermal and power-law components.
We suspect a similar bias in  a treatment of the underlying continuum in the
analysis of K$_\alpha$ line profile in the {\it Beppo}SAX data from the Black Hole 
Candidate source XTE J1650-500 by \cite{mfm04} who showed the red-skewed profile of the line and interpreted  as a broad and strongly 
relativistic iron emission line. However, recently \cite{mtf09} revisited  this particular set of
 {\it Beppo}SAX data from  XTE J1650-500 and reported
much weaker  K$_\alpha$ line emission when BMC is used for 
spectral continuum. We note that  BMC model, as a Generic Comptonization model, is more appropriate than empirical models based on the power law approximation of the non-thermal part of
the spectrum. Specifically, in application to low energy X-ray data the power-law component
predicts excessive low energy emission which then results in artificial broadening of the
iron line. BMC model, as well as COMPTT model for thermal Comptonization in case of NS, properly accounts for this low energy curvature and therefore is more relevant for the analysis of BH spectra.

In Appendix A we show that the approach used for a derivation of the Generic Comptonization Green's function (and for the Generic  Comptonization model as a  BMC) can be also applied to explain {\it an empirical  determined  exponential and power-law distribution of wealth and income}  in the United States and the United Kingdom.

 
 \section{Discussion on Possible Origin of  K$_\alpha$ Red-skewed Line Profiles in WD, NS and BH}
 
In this Paper we utilize   XMM-{\it Newton}/Epic-PN data
from accreting WDs, NSs and BHs to investigate in detail the structure 
 of a red-skewed line profile of iron K$_{\alpha}$  emission from these
sources. The asymmetric iron lines from NS and BH sources 
has been widely studied. The detections of the red-skewed lines in a number of Galactic and
extragalactic BHs have led to the conclusion that this particular
red skewness of the line in BH sources is related to General
Relativity (GR) redshift effects in the innermost part of BH
accretion disk \citep{mill07}. 
Moreover  the recent discovery of red-skewed iron line profiles in NS sources was also
 fashionably interpreted in terms of relativistic reflection model \citep{bs07,mill07,c07}.

However, we have to note that HM04 [see also \cite{mc07}] had
reported, using data from the Chandra HETG, the downshifted
shoulder  in  the K$_{\alpha}-$line profile of GK Per. 
Also \cite{v05} have  discovered a 
very strong red-skewed line profile at 0.6 keV  in  the  {\it Chandra} ASIC-S spectrum   of  GK Per.
They  also pointed out the  constancy of all observed  emission lines from GK Per including the Fe 
fluorescence line with spin period and they further suggest that  the emission site of these lines 
are visible at all times and from all viewing angles
during an orbit. This  Vrielmann's result on the constancy of the emission lines in GK Per provides one more 
evidence of the formation of the lines in the very extended configuration (presumably the wind) surrounding 
the source. These facts indicate an apparent similarity in observational appearances of red-skewed iron lines in 
WDs, NSs and BHs. However, the relativistic reflection paradigm does not offer a consistent explanation 
for the phenomenon across all types of objects. This motivated us to perform the presented
study of the iron lines in all three types of compact objects in order to test and compare 
conventional and alternative theories of the emission line formation.

We extract XMM-{\it Newton} spectra for WD source GK Per, NS source Ser X-1 
and BH source GX 339-4. We analyze the spectra with both relativistic models (DISKLINE and LAOR)
and the wind/outflow models (LT07). 
In Figures   \ref{GKPER_fit}-\ref{gx339_514_fit}  we present a comparison of  the
red-skewed line profile of GK Per  and  that observed in  NS Ser X-1  and BH GX 339-4
sources. While $\chi^2_{red}$ statistic value is slightly better for the WINDLINE model in the cases
of GK Per and GX 339-4, both models were able to successfully describe the 
line profiles in all types of objects. Thus  the   observations of the K$_\alpha$
emission with red-skewed features in CV GK Per suggest    that the
red skewness   of the line cannot be a signature of the GR effects, because the nature of the compact star in GK Per excludes any significant GR effect contribution. 
 If we assume the common origin of the line is WD, NS and BH then there is only one 
alternative related to the wind (jet) origin of the red-skewed line.
On the other hand one can argue that  the origin of the red-skewed K$_{\alpha}$ line in BH and WD sources is different given that in  BHs  the K$_{\alpha}$ line is  much more  red-skewed compared to that  in  WD (see upper  panels of Figs. \ref{GKPER_fit} and \ref{gx339_514_fit}).
The  natural question then arises: how one can use observations to choose between these two current 
paradigms of the origin of K$_\alpha$ line detected from X-ray sources?

In fact, variability properties can arguably provide means to identify 
conditions in the inner accretion flow and constrain the
physical processes governing the formation of the energy spectrum. Let us  compare  the shortest observable variability time scale seen in PDS of GK Per  with a theoretical estimate  for the WD emission region.  In Figure \ref{pds_gkper} we present the XMM  power spectrum of GK Per related to the XMM photon spectrum  shown in  Fig. \ref{GKPER_fit}.  One can see from this Figure that the variability   with time scales shorter than 30 s (or in terms of frequency higher than 0.03 Hz)  is not present  in the data.  Provided that the shortest dynamical characteristic time scale 
$T_{ch}\sim \nu_{ch}^{-1}$ is determined by a ratio of size  of WD  emission region $L_{WD}$ and acoustic velocity $V_{A}$ we obtain  that  $T_{ch}\sim 30 ~(L_{WD}/10^{9}{\rm~cm})/(V_{A}/3\times10^7{\rm~cm})$ s. 
Here we assume that $L_{WD}$ is of order of WD radius, i.e $10^{9}$ cm  and $V_{A}\sim 10^{-3}\times 
[kT({\rm keV})]^{1/2}c$. Note the best-fit plasma temperature kT for GK Per spectra is about  a few keV (see Table \ref{gkper_tab})
and thus  our theoretical expectations perfectly agree with the data. 
The highest frequency seen in PDS for GK Per is
0.03 Hz ($\sim1/T_{ch}$).  Specifically, the variability 
shorter than the dynamical time scale (as a ratio $L/V_{A}$) is not seen in PDS. In other words
{\it the highest frequency is seen in PDS is directly related to the shortest dynamical scale of the emission region}.

M04c presented a number of arguments against the wind/outflow 
mechanism of the line production. First, M04c presumed that high
optical depth of the wind $\tau_{W}\gg1$ can prevent us from seeing
the BH detected high QPO frequency (of a few times 100 Hz). In fact,
in the observation of BH GX 339-4 we do  find  a  value
of $\tau_{W}$ of 4.7. M04c further noted that line emission unaffected by the relativistic
regime near BH must occur  at radii $R_{W}$ greater than 100 $R_{\rm
S}$ thus  M04c argued that the wind model requires the mass outflow
rate $\dot M_{of}$ to exceed the mass inflow rate $\dot M_{d}$  by
at least factor a few. Using the best-fit parameter values of
$\tau_{W}$ and $\beta_W$ for GX 339-4 we find that the inferred $\dot
M_{of}\sim 6 \times10^{18}A(\tau_{W}/4.7)(R_{W}/100R_{\rm
S})(\beta/0.0.026) $ g s$^{-1}$. Here $A= S_{W}/4\pi R_{W}^2\leq1$ is
the dimensionless cross section of the wind and Schwarzschild radius $R_{\rm S}=3\times
10^5m$ cm is calculated for 10 solar mass black hole (i.e.
$m=M/M_{\odot}=10$).
In contrast to the M04c's claim we find that, taking into account  uncertainties 
of parameters entering in $\dot M_{out}$ and 
$\dot M_{d}$,  the mass outflow rate 
$\dot M_{out}\lax \dot M_{d}$ 
given that in the high/soft state of BH   $\dot M_{d}\sim
3\times 10^{18}(l/0.2) (m/10)/(\varepsilon/0.1)$ g s$^{-1}$ where
$l=L/L_{\rm Edd}$ is the Eddington ratio and   $\varepsilon=L/(\dot
M_{d}c^2)$ is  an efficiency to convert the mass accretion rate to
X-ray radiation. 
Thus,  we show that arguments used by M04c to falsify
the wind downscattering scenario for iron line production in BH sources
do not hold for the presented data from GX 339-4.

Furthermore, according to the   M04c's arguments based on the relativistic model, 
if the GR origin of the line is to prevail for the BH case then the   
observations of  the strongly  red-skewed K$_{\alpha}$ should be accompanied  by the detection of short time scale variability in the form of QPO frequencies in the range of 1-10 Hz and higher  or 
broad band Fourier power density  components (or even both), 
because both of them are expected to originate in the innermost part of the emission region.

However, the observational situation in GX 339 is in contradiction to this 
consequence of the relativistic model.
When the strong K$_{\alpha}-$line was detected (Observation 1 and 2),
the simultaneous observations by {\it RXTE}/PCA showed a typical Comptonization spectra  of the high/intermediate soft state with photon 
index $\Gamma\sim 2.4-2.6$  (see Table \ref{gx339_tab}) along
with  a low variability, noisy (forest-type) PDS
 with no QPO features and no broken power-law continuum 
 (see Fig. \ref{gx339_ga_eufs}, {\em right upper and middle panels}).  
On the other hand, during the XMM observation on 2003 March 8 (Observation 3) when the iron line was 
weak  the {\it RXTE} observations showed a PDS with  a clean 
feature of  a broken power-law (white-red noise component) with a break frequency at about 2 Hz  
and QPOs with centroid frequencies of 8 and 16 Hz. In the next two subsections we show that 
the suppressed high frequency variability can be readily explained by the presence 
of the optically thick wind outflow.

As a result   we find that both relativistic and wind/outflow model are able to successfully  fit 
the data on iron emission lines from compact X-ray sources. Even formally applied for the 
WD case relativistic models satisfactory fit the data. Therefore,
red-shift itself is not a reliable signature of the GR effects. Relativistic reflection 
scenario cannot explain the lack of variability of the line observed in CVs \citep{v05}
and is in contradiction with the observation of the suppressed high frequency variability 
during the ``line''-state in BH source GX 339-4. Hence, considered in the context
of variability properties the wind/outflow scenario emerges as a favorable explanation 
of red-skewed lines in accreting X-ray binaries.

\subsection{The wind effect on the resulting power spectrum}
In the presence of the optically thick outflow the emergent timing signal $W(t)$ is  a convolution   
of X-ray timing signal originated in  the innermost part of the source $X(t)$  and the outflow 
transformation function   due  to photon scattering $\varphi(t)$ see \citep[see][]{t07}:
\begin{equation}
W(t)=\int_0^t X(t^{\prime})\varphi(t-t^{\prime})dt^{\prime}.
\label{eq:emerg_signal}
\end{equation}
The  response of any bounded configuration   due to photon scattering is exponential 
[see e.g.   \cite{st80} and \cite{lt07}] . Thus 

\begin{equation}
\varphi(t)=A\exp(-t/t_0)
\label{eq:outflow:response}
\end{equation}
where $t_0\sim\tau_W L_W/c$ is the outflow (wind)  scattering (photon diffusion) time scale,
$\tau_{W}$ is the outflow Thomson optical depth, $L_{W}$ is the scattering wind size and $A=1/t_0$ is the
 normalization constant ($\int_0^{\infty} X(t)dt=1$).    

The resulting power spectrum $||F_W(\omega)||^2$ is a product of the power spectra of $\varphi(t)$ and $X(t)$, namely
\begin{equation}
||F_W(\omega)||^2=||F_X(\omega)||^2||F_{\varphi}(\omega)||^2
\label{eq:res_pds}
\end{equation} 
where $F_W(\omega), ~F_X(\omega), ~F_{\varphi}(\omega)$ are Fourier transforms of 
$W(t), ~X(t),~\varphi(t)$ respectively, and $\omega=2\pi\nu$ is a rotational frequency.

The power spectrum of $\varphi(t)$ is a centered Lorentzian 
\begin{equation}
||F_{\varphi}(\omega)||^2=\frac{(1/t_0)^2}{\omega^2+(1/t_0)^2}.
\label{eq:varphi_pds}
\end{equation}

If for example,   X-ray signal $X(t)$ can be presented  a quasi-periodic (QPO) oscillations related 
to the frequency $\omega_{\ast}=2\pi\nu_{\ast}$ then the corresponding PDS is a Lorentzian of the form:
\begin{equation}
||F_{X}(\omega)||^2=\frac{C_N}{(\omega-\omega_{\ast}^2)+(\hat \Gamma/2)^2}.
\label{eq:X_pds}
\end{equation}
where $\hat\Gamma$ is a FWHM and $C_{N}$ is the normalization constant of the QPO PDS.   

In Figure \ref{photon_scat_pds} we present the plot of  power spectra  $||F_{X}(2\pi\nu)||^2$ and 
$||F_{\varphi}(2\pi\nu)||^2$ along with the resulting PDS\ $||F_W(2\pi\nu)||^2$. 
One can see from upper panel of Fig.  \ref{photon_scat_pds} that when $\nu_{\ast}$ and
$1/(2\pi t_0)$, the characteristic frequencies 
of X-PDS and $\varphi-$PDS  respectively [or characteristic time scales $1/(2\pi \nu_{\ast})$ and $t_0$], are comparable 
then the PDS intrinsic features of  X-ray signal originated in  the innermost part of the source is 
unaffected by photon scattering in the wind.  In the opposite case when  $\nu_{\ast}>>1/(2\pi t_0)$
the QPO feature of X-PDS is  completely washed out, see black line in lower panel of  
Fig.  \ref{photon_scat_pds}.

The similar washing out effect can be seen in the general case when  X-PDS is presented a superposition of 
white-red noise (broken power law)   and QPO Lorentzians. \citep[see e.g.][]{t07}.   
All power  related to frequencies $\nu\gg1/(2\pi t_0)$ are suppressed due photon scattering in the wind. 

\subsection{Forest type of PDS in the "line" state}

The observational correlation of the strong  line and the forest-type  power spectrum gives a 
strong support for the wind (outflow) origin of the  line in BH at least in GX 339-4.  For a particular 
source the presence of a strongly red-skewed iron line along with the forest-type PDS in the 
data can be considered an observational signature of wind (or wide open jet) [see above cases of GK Per and GX 339-4, (Observation  2)]. 

Namely, when the powerful wind occurs, presumably  as a consequence of high mass rate in the source, 
the wind  is illuminated by the emission of the innermost part of  BH source to 
produce the highly red-skewed iron line but  on the other hand any variability of X-ray innermost emission region is suppressed by the wind given its extension. 
For the characteristic size of the wind $L_{W}\gax 10^3 R_{\rm S}$, the wind optical depth of 4.7 
and 10 solar masses BH, i.e. $m=M_{BH}/M_{\odot}=10$,  the photon scattering time in the wind  
(see Eq. \ref{eq:outflow:response})
\begin{equation}
t_0\gax 0.5\left( \frac{\tau_{W}}{4.7}\right)\left[\frac{L_{W}}{10^3 (m/10)(R_{\rm S}/3\times10^6~{\rm cm})}\right]\frac{3\times 10^{10}~{\rm cm}~{\rm s^{-1}}}{c} 
~{\rm s}.
\label{wind_scat_time}
\end{equation}  
Thus according to the above inequality all frequencies higher than
\begin{equation}
\nu>1/(2\pi t_0)\sim 0.3\left( \frac{\tau_{W}}{4.7}\right)^{-1}\left[\frac{L_{W}}{10^3 (m/10)
(R_{\rm S}/3\times10^6~{\rm cm})}\right]^{-1}\left(\frac{3\times 10^{10}~{\rm cm}~{\rm s^{-1}}}{c}\right)^{-1}~{\rm Hz} 
\label{qpo_freq_smear}
\end{equation} 
are smeared  (suppressed) due the photon scattering in the wind. This effect of the wind photon scattering is presumably seen in the PDS in a  so called ``line state" when the strong  line is observed  
in Observations 1 and 2 of BH GX 339-4 (see middle and top  panels of  Fig. \ref{gx339_ga_eufs}).

 \subsection{Variability of X-ray emission, line appearances and so called   ``reflection'' effect}
As we show above the detection  of the strong K$_{\alpha}$ line along with a weak  variability 
of X-ray can be an observational signature of presence of the powerful wind in the system
(see Fig. \ref{gx339_ga_eufs}, upper and middle panels). 
On the other hand one  may  also  expect  the weak line along with  high variability of  X-ray emission of the source as a signature of the weak wind (see Fig. \ref{gx339_ga_eufs}, low panels). 
Thus  a noticeable fraction of X-ray  variability presumably  originated in the
innermost part of the source $\exp(-\tau_W )$ is unaffected by the wind and passes  through the wind
 if $\tau_W\lax 1$.      Indeed, in Observation 3 we probably see an optically thin wind  which signatures are a weak line having almost Gaussian profile and noticeable high variability.   
LT07 considered only  the line scattered component  
because the line profile of the direct component  is very sensitive to the wind geometry and to the position of the Earth observer.  
\subsubsection{Formation of  direct component of the wind line}
 
 The direct component of the  windline model  may be fitted by  the Gaussian line profile if the wind  is symmetric with respect to the Earth observer (see an example in Appendix B). In the opposite case   the line profile can be asymmetric, blue-skewed or red-skewed.
The wind line profile of  the optically thin wind  is described by the direct component  only.  In this case  strong variability of the central engine is  unaffected because the attenuation of this signal is  
 $\exp(-\tau_W)\sim 1$.  
LT07 showed that  the source of K$_{\alpha}$  in the wind is mostly located in the narrow shell near the innermost part of the wind (see details in LT04 and
LT07).
 One can assume that K$_{\alpha}$ photons are emitted isotropically by a given element  of the wind shell of  radius $r$     and the local line profile is described by Gaussian
\begin{equation}
I_L(E,r,\mu)=\frac{1}{\pi^{1/2}\Delta E_{\rm D}}\exp\left\{-\frac{[E-E_0(1+\beta_W(r)\mu)]^2}{\Delta E_{\rm D}}\right\}
\label{local_line_prof}
\end{equation} 
where   $\Delta E_{\rm D}=E_0(kT_W/m_{Fe})^{1/2}/c$ is the intrinsic Doppler width of the line which is very small, i.e.  
$\Delta E_{\rm D}=1.3\times 10^{-4}(kT_W/1~{\rm keV})^{1/2}E_0$,  $\mu=\cos{\theta}$
and  $\theta$ is angle between  the velocity vector ${\bf V}_{W}$  and 
the  vector  the line of sight ${\bf n}_{ert}$,
 i.e $\mu=({\bf n}_{W}\cdot {\bf n}_{ert}$), 
${\bf n}_{W}={\bf V}_{W}/V_{W}$ is  unit vector of  ${\bf V}_{W}$.
This line profile can be derived using the bulk local Maxwellian  distribution for  atom (iron) velocities 
\citep{tz98} and Doppler effect [see \cite{rl79} for details of this derivation for the pure thermal motion].   

The line flux  emitted by the wind shell  detected by the Earth observed $F_L(E)$ is  a sum (integral)  of a projection of the line intensity  of each element of  the wind shell to the line of sight. Namely,
$$
F_L(E) =\frac{1}{E_0}\int_{R_{in}}^{\infty}\frac{F(r)r^2}{\beta_W(r)}dr \int_0^{2\pi}d\varphi\times
$$
\begin{equation}
[\beta_W(r)E_0]\left[\int_{\mu_{1,+}}^{\mu_{2,+}}|\mu| 
I_L(E, r,\mu)d\mu^{\prime} +\int_{\mu_{2,-}}^{\mu_{1,-}}|\mu| 
I_L(E, r,\mu)d\mu^{\prime}\right]
\label{line_flux}
\end{equation}
where $F(r)$ is the line emissivity at $r$, $\mu={\bf n}_{W}\cdot {\bf n}_{ert}=\mu^\prime\mu_0+(1-\mu^{\prime 2})^{1/2}(1-\mu_0^2)^{1/2}\cos(\varphi-\varphi_0)$, $ {\bf n}_{W}= \{(1-\mu^{\prime2})^{1/2}\cos\varphi, (1-\mu^{\prime2})^{1/2}
\sin\varphi, \mu^\prime\}$, $ {\bf n}_{ert}= \{(1-\mu_0^2)^{1/2}\cos\varphi_0, (1-\mu_0^2)^{1/2}
\sin\varphi_0, \mu_0\}$ and  $\mu_{1,+}$,  $\mu_{2,+}$ and  $\mu_{1,-}$,  $\mu_{2,-}$  are cosines of angles of toward and  outward wind cones, for which ${\bf n}_{W}\cdot {\bf n}_{ert}>0$ and 
${\bf n}_{W}\cdot {\bf n}_{ert}<0$   respectively.
For  the  symmetric case   $\mu=\mu^{\prime}$,  $-\mu_{1,-}=\mu_{1,+}\geq0$ and 
$-\mu_{2,-}=\mu_{2,+}>0$ .

In Appendix B  we show  how one can obtain the Gaussian-like  profile of the wind line 
from  the spherical wind (see Fig. \ref{wind_geom}).  The  general case of the wind geometry and relative position of the Earth observer with respect to the wind  will be presented elsewhere. 
 The Earth observer could detect  the line profile 
if she/he  could spatially  resolve a contribution of  each particular spherical shell of the wind. 
 We can use  the steepest descend method, given that   $\Delta E_{\rm D}\ll E_0$,  to calculate the internal  integral $I_{in}$ of Eq. (\ref{line_flux}). 
 We obtain that the profile of the line coming from a given spherical shell (see Fig. \ref{wind_geom})
\begin{equation}
{\cal I}(E, r) =[\beta_W(r)E_0]I_{in}=
\left\{ \begin{array}{ll}
\frac{|E-E_0|}{\beta_W(r) E_0}
  & \mbox{ for $|E-E_0| \leq \beta_W(r) E_0$;}\\ 
 0  & \mbox{for $|E-E_0| \geq \beta_W(r) E_0 $.}
\end{array}
\right.
\label{rad_flux}
\end{equation}
is a broken power-law with the center at $E=E_0$. 
The resulting direct  component of the line profile is a sum of the contribution of each of the shells, i.e. the sum of the product of the line emissivity and  the broken power-line profile ${\cal I}(E, r)$
(see Eq. \ref{res_flux}). 
Note if the wind is not symmetric with respect to the observer then the line profile  is also asymmetric with respect to $E_0$ (the line rest-frame energy). 

Thus one can see that  the line  profile can  be  presented by the Gaussian-like shape in the case of optically thin symmetric wind.  
Moreover we show  in Appendix B that the inferred Gaussian width $\sigma_{ga}$
can be related to the mean dimensionless  bulk velocity 
$\beta_{mn}\sim\beta_{W}\sim\sigma_{ga}/ E_{0}$.  
We can infer the wind velocity $V_W\sim0.1$c for Observation 3 using this relation for $\beta_{W}$
 and the inferred values of $\sigma_{ga}\sim0.69$ keV and  $E_{L}=E_0\sim 6.4$ keV (see Table 3).  The wind velocity 0.1c  is consistent with an optically  thin wind in the case of  Observation 3.  
Indeed if the mass outflow does not change   through Observations 1-3 one can estimate, using the continuity equation  for the wind, that  $\tau_W$ should be inversely proportional to $V_W$ and then
 $\tau_{W,3}\sim( V_{W,2}/ V_{W,3}) \tau_{W,2}\sim1$.
Such a low  value  of $\tau_{W,3}$ is consistent with the presence of high variability in Observation 3 because, at least,  $1/3$ of variability of the central engine can be  directly  seen and unaffected by the wind.
\subsubsection{On the ``relativistic'' line picture}
One can ask a fair question: could  be that the wind line model holds in some cases (e.g., High Soft State and Soft Intermediate State), whereas the relativistic line picture holds elsewhere (Hard Intermediate State)? The suggested   relativistic line is a combined effect  of the presence of the cold target (disk) in the innermost part of the source and high visibility of the source  innermost part.  
The high visibility  is, indeed, confirmed by some particular  observations of BH sources when high frequencies QPOs are seen in their  power spectra up to 200 Hz [see \cite{rm}]. 
However  the presence of cold unionized material in the innermost part of the source is strongly ruled out by ionization state of this region.   LT04 demonstrated that the innermost part of the disk is completely ionized. They used  extensive calculations of interaction of  X-ray hard radiation emanated from the central region  with surrounding plasma.  

The relativistic line paradigm  is also related to  the reflection problem  of  X-ray radiation.
In fact, in our fits of the GX 339-4 spectra, we  find essentially no need for reflection 
but it is not by chance that the reflection  bump is not seen in the data.
We remind a reader that the reflection bump is a combined effect of downscattering hard photons of the incident spectrum and photoelectric absorption of photons which  energies  are lower than 10 keV \citep{bst74}. It is impossible to form a downscattering bump when the spectral continuum is very soft,  like in the GX 339-4 spectra presented here, for which photon  indices $\Gamma=\alpha +1$ are about 2.5 (see Table \ref{gx339_sept292002_tab}).  In fact, LT07 demonstrated using analytical consideration and Monte Carlo simulations that for  $\Gamma>2$  there is {\it no}  bump in the reflection spectrum,  due to downscattering accumulation of photons from high energy tail of the incident spectrum.  In other words the incident spectrum  is too steep in order to have enough photons to form the downscattering  bump in the reflection spectrum.  
On the other hand as we argue above  the innermost part of the disk is  completely ionized and there is no ``cold'' disk there. Note  the subtended angle of  the hard X-ray  emission emanated from the source central part  is highest for the disk located in the same central area.  
Therefore we can conclude  that the reflection bump cannot be formed,  in the innermost part of the disk,  as a  result of the photo-electric absorption and  photon dowscattering  of hard X-ray photons.   

\section{Conclusions}

We utilize XMM-{\it Newton} data from Epic-PN Camera to study the red-skewed iron lines 
in GK Per, Ser X-1 and GX 339-4 as representatives of 
WD, NS and BH X-ray binary sources respectively. We analyze the iron line profiles in terms of both 
relativistic reflection and wind outflow models. We also use simultaneous 
{\it RXTE} data to obtain information on high energy part of the spectrum and 
high frequency variability for GX 339-4.

We base our conclusions on the following facts: i. all three types of
accretion compact objects show red-skewed iron $K_\alpha$ lines, ii.
outflow is a common phenomenon for CVs, NSs and BHs,    iii. we demonstrate the
ability for the wind model to describe the
asymmetric shapes of the lines  detected from CV, NS and BH
sources and iv.  we show  the  observational correlation of the strong  line and 
forest-type   power spectrum in BH. The lack 
of high frequency variability weakens the red-skewed line connection with high frequency QPOs
and strengthens its connection to a wind/outflow phenomena.
We should also  point out whereas a lack of high frequency variability could be counted as observational support for the wind model however it cannot be counted as observational evidence against the relativistic line model. At least without knowing what the underlying variability spectrum should be.
Moreover, we also point out that Gaussian-like profiles of relatively narrow iron lines, $\sigma_{ga}\sim \beta_WE_L\ll E_L$, can be a direct line component coming from optically thin wind.  

These points and the difficulty for the GR effects to account for
red wings in the iron lines observed in CV sources
favor a wind origin of the red-skewed  line profile of
K$_{\alpha}$ iron emission in all these sources.

The authors acknowledge the productive discussion the paper content with 
Martin Laming, Michael Wolff, Kent  Wood, Marat Gilfanov,  Filippo Frontera, Remo Ruffini, Vadim  Arefiev, Elena Seifina and Mike Nowak. We are very grateful to the referee  for  his  useful comments and suggestions which significantly  improve the paper presentation.

\appendix
\section{The Comptonization  Green's function as a Generic Green's function for photon upscattering 
 off energetic  electrons of plasma cloud and for capital gain due to investment}
\cite{dy01}, hereafter DY01,  present the data   on wealth and income distributions in the United Kingdom (UK) , as well as on the income distributions in the individual states of the USA. They found  that the great majority of wealth and income distribution (95\%) is described by an exponential distribution, whereas the high-end tail follows a power law  (see Fig. 1-3 in DY01).
In fact, the exponential gross income distribution is similar to blackbody (BB)-like distribution 
$B_{ph}(E)$ for photons emerged from the optical thick configurations such either a disk or a stellar atmosphere. Actually 
to fit the data DY01 used  the exponential law $P_{maj}(w) \propto \exp(-w/W)$ where $P(w)$ stands for  probability density to have a wealth ${\cal W}$ between $w$ and $w+dw$  and $W$  is a characteristic (mean) wealth  for a given data selection (for either population of USA or UK). 
The mean wealth $W$ has a full analogy with BB temperature.  The exponential law, also known in physics as the Boltzmann$-$Gibbs distribution, is characteristic for a conserved variable, such as
energy. DY01  argued that, because money (cash) is conserved, the probability
distribution of money should be exponential. Wealth can increase or decrease
by itself, but money can only be transferred from one agent to another. So, wealth is
not conserved, whereas money is.

Below we show the photon distribution due upscattering of soft (seed)  photons off energetic electrons 
as a power law is similar to the capital gain distribution due to investment, for example in a stock market.
We denote the intensity 
of the injected monochromatic soft photons of energy $E_0$ and  gaining  energy  due to scattering  in the Compton cloud as  $I(E_0,E)$ and $N_{ph}=I(E_0,E)/E$.  
As for the  capital gain  due to investment $M(w_0,w)$ and $P(w_0,w)= M(w_0,w)/w$. 
Because the derivation is similar for photon  and  wealth distributions we present this for  $N_{ph}(E)$ and  through the derivation we draw  a parallel between terminology for photons and capital gains.  
Note that in the system, disk+Compton Cloud  only some fraction of  seed  (disk) photons $f$ is affected by upscattering off CC energetic electrons but (1-f) fraction of seed  photons comes directly to the observer. The same is true for the wealth distribution.  Only some fraction of money  $f$ belonging to 
$N_{maj}(w)-$part is invested into either stock market or big business.  The (1-f)-fraction of the majority of population does not contribute into  this business activity. 

The intensity (amount of capital gain) of the injected soft photons of energy $E_0$ 
(or investment of amount $w_0$)   undergoing $k$ scatterings  in the Compton 
cloud is 
\begin{equation}
I(E_0, k)\propto p^k
\label{int_k}
\end{equation}
where $p$ is  average probability of photon scattering in the CC , or in the case of capital gain is average probability of success  due to investment.
Note that as in the photon case 
the probability of  photon scattering $p$ is directly related to  mean number of scatterings (or successful investments)
\begin{equation}
{\cal N}_{sc}=\sum_{k=1}^{\infty}kp^kq=p/(1-p)
\label{scat_number}
\end{equation}
where $q=1-p$ is the probability of the photon escape from the CC or that of the failure in the investment case.  
Thus, using Eq. (\ref{scat_number}), we obtain 
\begin{equation}
p=1-1/({\cal N}_{sc}+1).
\label{p_vs_sc_number}
\end{equation}
Because the average photon energy (money) change   per scattering  (investment) $<\Delta E> =\eta E$ 
(where $\eta>0$ for the upscattering case or successful investment),  
the injected photon energy after $k$ scatterings  $E$ is
\begin{equation}
E=(1+\eta)^{k}E_0 
\label{E_k}
\end{equation} 
For simplicity of presentation here we drop subscript $k$ for the value of photon energy $ E_k$ after k-scattering (or for capital gain after k successful investments). 
 
The combination of Eqs. (\ref{int_k}),  (\ref{E_k})  yields that the emergent upscattering spectrum of the soft photon of energy $E_0$ in the bounded Compton cloud is a power law
\begin{equation}
I (E_0, E) \propto\left(\frac{E}{E_0}\right)^{-\alpha}
\label{power_law}
\end{equation}
 which  index 
\begin{equation}
\alpha=\frac{\ln(1/p)}{\ln(1+\eta)}.
\label{alpha_pl}
\end{equation}
 Using Eq. (\ref{p_vs_sc_number})
  we can reduce Eq. (\ref{alpha_pl})  to 
\begin{equation}
\alpha\approx(\eta {\cal N}_{sc})^{-1}=Y^{-1}.
\label{alpha_plm}
\end{equation}
for ${\cal N}_{sc}\gg1$ and $\eta\ll 1$.
Thus the photon  distribution  over E [or probability of the capital  distribution of the population over $w$,
$N(w_0,w)$] 
is 
\begin{equation}
N_{ph}(E_0, E)= I (E_0, E)/E \propto\left(\frac{E}{E_0}\right)^{-\Gamma}
\label{power_law_ph}
\end{equation}
where $\Gamma=\alpha+1$.
In order to obtain the upscattering  (capital gain) component of the resulting photon (capital) distribution  $N_{ph, res}(E)$ 
we should convolve the upscattering photon Green's funcition   $N_{ph}(E_0, E)$ 
[or $N(w_0, w)$] with some fraction $f$ of a seed photon  distribution  $B_{ph}(E_0)$
 [or with some fraction $f$ of the  majority exponential capital distribution $P_{maj}(w_0$) in the case of gross income distribution]. Then we add the direct component of the seed photons unaffected by upscattering $(1-f)B(E)$ with the upscattering photon component, i.e. 
 \begin{equation}
N_{ph, res}(E)=(1-f)B_{ph}(E) +f\int_0^\infty N_{ph}(E_0, E)B_{ph}(E_0)dE_0.
 \label{emerg_phsp}
 \end{equation}
It is easy to show that the shape of  the distribution $N_{ph,res}(E)$ [or $N_{res}(w)$] is an exponential  up to some energy 
and then it follows by power law to high-end of the spectrum.
Thus we can conclude that the photon Comptonization  spectrum  as a convolution of 
the Green's function (as a broken power law) with the seed photon spectrum (BMC model) is applicable to describe the resulting shape of the  USA and UK measured wealth distributions. 
\section{The direct line component}  
 
 As it follows from Eqs. (\ref{line_flux}-\ref{rad_flux}) the resulting direct  component of the line profile is 
 \begin{equation}
F_L(E) =2\pi x \int_{R_{min}}^{\infty}\frac{F(r) r^2 dr}{\beta_W^2(r)} 
\label{res_flux}
\end{equation}
where $x=|E-E_0|/E_0$ and $R_{min}$ ($> R_{in}$) is determined by condition that
\begin{equation}
\frac{xf(R_{min})}{\beta_W^2(R_{min})}= \frac{f(R_{min})}{\beta_W(R_{min})}
\label{cond_rmin}
\end{equation}
which identical to 
\begin{equation}
x=\beta_W(R_{min}).
\label{cond_rmin_mod}
\end{equation}
This equation can be solved analytically if  we assume that
\begin{equation}
\beta_W(r)=a (r-R_{in})^{\alpha}
\label{wind_vel_power-law}
\end{equation} 
where $\alpha>0$. 

We can introduce a new integration variable $y=R-R_{in}$ and rewrite Eq. (\ref{res_flux}) using 
Eq. (\ref{wind_vel_power-law}) as follows 
 \begin{equation}
F_L(E) =2\pi \frac{x}{a} \int_{(x/a)^{1/\alpha}}^{\infty}\frac{F(R_{in}+y) (R_{in}+y)^2}
{\beta_W(y)}\frac{dy}{y^{\alpha}} 
\label{res_flux_m}
\end{equation}

Because the line emissivity decays exponentially with respect to $y$ (see LT04 and LT07)
we can also assume that
\begin{equation}
{\cal W} (y)=\frac{(R_{in}+y)^2F(R_{in}+y)}{\beta_W(y)} = C_0(R_{in}) \frac{1}{y}\exp(-ba^2y^{2\alpha})
\label{F_R2_cond}
\end{equation}
where $b>0$.
With this  assumption regarding function $F(r)$  the integral in Eq. (\ref{res_flux_m}) diverges  at the lower  limit of integration and thus we obtain the Gaussian-like profile.
 \begin{equation}
F_L(E) \propto \exp(-bx^2).
\label{res_flux_mm}
\end{equation}
One can ask a natural question how the coefficient $b$ of  exponential of Eq. (\ref{F_R2_cond}) is related  to the mean value of the dimensionless  bulk velocity $\beta_{mn}$ which  by definition  is 
\begin{equation}
 \displaystyle{
\beta_{mn}=\frac{\int_0^{\infty} \beta_W(y){\cal F} (y)(y)dy}{\int_0^{\infty}{\cal F} (y)dy}.
 }
 \label{Vmn}
\end{equation}
where 
${\cal F}(y)= \beta_W(y){\cal W}(y)$. 
Using Eqs. (\ref{wind_vel_power-law}, \ref{F_R2_cond}) we can rewrite Eq. (\ref{Vmn}) as follows:
\begin{equation}
 \displaystyle{
\beta_{mn}=\frac{a^2\int_0^{\infty} y^{2\alpha-1}\exp(-ba^2y^{2\alpha})dy}
{a\int_0^{\infty} y^{\alpha-1}\exp(-ba^2y^{2\alpha})dy}.
 }
 \label{Vmn_mod}
\end{equation}
To calculate integrals in Eq. (\ref{Vmn_mod}) we introduce new integration variables 
$t=a^2b y^{2\alpha}$  and $z=ab^{1/2} y^{\alpha}$, for the numerator and denominator respectively.
Then we obtain that 
\begin{equation}
\beta_{mn}=D_0 b^{-1/2}
 \label{Vmn_final}
\end{equation}
where the numerical factor 
\begin{equation}
D_0=\frac{\alpha-1}{\pi^{1/2}(\alpha-1/2)} \lax 1
\label{D_0}
\end{equation}
is of order of one if $\alpha>1$.

\clearpage
\newpage

\begin{deluxetable}{lllll}
\tablewidth{0pt}
\tabletypesize{\footnotesize}
\tablecaption{Best-fit Parameters for the {\it XMM-Newton} Spectrum of CV GK Per}
\tablehead{
\colhead{Component} &
\colhead{Parameter} &
\colhead{LAOR} &
\colhead{DISKLINE} &
\colhead{WINDLINE}}
\startdata
EDGE& $E_{edge}$ (keV) & 7.11$\pm$0.02 & 7.09$\pm$0.03 &7.09$\pm$0.04 \\
& $\tau_{max}$ & 0.35$\pm$0.03  & 0.36$\pm$0.02& 0.23$\pm$0.03 \\
WABS& $N_H$ (cm$^2$)   & 1.5$\pm$0.3 & 1.2$\pm$0.1 & 2.0$\pm$1.0 \\
BBODY& $T_{bb}$, keV   & 0.16$\pm$0.04 & 0.15$\pm$0.01 & 0.12$\pm$0.01 \\
COMPTT& $T_0$ (keV)  & 0.9$_{-0.1}^{+0.5}$ & 1.18$\pm$0.01 & 1.05$\pm$0.57 \\
& $kT_e$  (keV)  & 3.40$\pm$0.07 & 3.4$\pm$0.3 & 5.3$\pm$0.5 \\
& $\tau_e$  & 32.3$\pm$4.5 & 27.0$\pm$0.2 & 20.9$\pm$1.8 \\
K$_\alpha$ line\tablenotemark{a} &$E_L$ (keV) & 6.49$\pm$0.02 & 6.46$\pm$0.03 & 6.48$\pm$0.05 \\
&$\beta_L$  & 2.14$\pm$0.16 & 2.4$\pm$0.2 & - \\
&$R_{in}$ ($R_G$) & 12.5$\pm$2.5 & 22$\pm$3 & - \\
&$R_{out}$ ($R_G$) & 400\tablenotemark{f} & 1000\tablenotemark{f} & - \\
&$i_L$ (deg) & 4$\pm$4 & 13.5$\pm$1.5 & - \\
&$\tau_W$ & - & - & 1.46$\pm$0.06 \\
&$\beta_{W}=V_{W}/c$ & - & - & (0.71$\pm$0.06)$\times 10^{-2}$ \\
&$kT_W$ (keV)  & - & - & 0.97$\pm$0.06 \\
&$EW_L$, eV..........  & 197$\pm$15 & 176$\pm$30 & 641$\pm$90 \\
\hline
Fit quality &$\chi^2_{red}$ ($\chi^2$/$N_{dof}$) & 1.19 (1114.6/934) & 1.20 (1123/934) & 1.15 (1071/934) \\
\enddata
\tablenotetext{a}{Model is specified by the name at the top of a column} 
\tablenotetext{f}{Parameter fixed}
\label{gkper_tab}
\end{deluxetable}

\newpage
\begin{deluxetable}{lllll}
\tablewidth{0pt}
\tabletypesize{\footnotesize}
\tablecaption{Best-fit Parameters for the {\it XMM-Newton} Spectrum of NS Serpens X-1}
\tablehead{
\colhead{Component} &
\colhead{Parameter} &
\colhead{LAOR\tablenotemark{a} } &
\colhead{DISKLINE} &
\colhead{WINDLINE}}
\startdata
WABS&$N_H$ (cm$^2$) & 0.32$\pm$0.02 & 0.35$\pm$0.01 & 0.37$\pm$0.02 \\
BBODY&$T_{bb}$ (keV) & 0.29$\pm$0.01 & 0.27$\pm$0.01 & 0.25$\pm$0.01 \\
COMPTT&$T_0$ (keV) & 0.55$\pm$0.02 & 0.518$\pm$0.002 & 0.50$\pm$0.02 \\
&$kT_e$ (keV) & 3.01$\pm$0.01 & 2.91$\pm$0.04 & 2.91$\pm$0.05 \\
&$\tau_e$  & 13.1$\pm$0.1 & 13.4$\pm$0.1 & 13.4$\pm$0.2 \\
K$_\alpha$ line&$E_L$ (keV) & 6.57$\pm$0.02 & 6.57$\pm$0.04 & 6.67$\pm$0.01 \\
&$\beta_L$ & 1.67$\pm$0.16 & 2.12$\pm$0.08 & - \\
&$R_{in}$ ($R_G$)  & 1.23\tablenotemark{b} & 6.0\tablenotemark{b} & - \\
&$R_{out}$ ($R_G$)  & 400\tablenotemark{f} & 1000\tablenotemark{f} & - \\
&$i_L$ (deg)  & 65$\pm$3 & 62$\pm$8 & - \\
&$\tau_W$ & - & - & 1.6$\pm$0.1 \\
&$\beta_{W}=V_{W}/c$  & - & - & (1.5$\pm$0.1)$\times 10^{-2}$ \\
&$kT_W$ (keV) & - & - & 1.21$\pm$0.06 \\
\hline
&$\chi^2_{red}$ ($\chi^2$/$N_{dof}$) & 1.15 (2292/1991) & 1.13 (2245/1991) & 1.14 (2264/1991)  \\
\enddata
\tablenotetext{a}{Model is specified by the name at the top of a column} 
\tablenotetext{b}{Parameter value pegged into the lower limit set by the model} 
\tablenotetext{f}{Parameter fixed}
\label{serpx1_tab}
\end{deluxetable}

\newpage
\begin{deluxetable}{llllllll}
\tablewidth{0pt}
\tabletypesize{\footnotesize}
\tablecaption{Gaussian Fit for the Iron Line for GX 399-4 observations in Different BH States}
\tablehead{
\colhead{Obs.}&
\colhead{Date}&
\colhead{BH state\tablenotemark{a}} &
\colhead{$\Gamma$ ($\alpha_{BMC}+1$)} &
\colhead{$f_C$\tablenotemark{b}} &
\colhead{$\sigma_{ga}$ (keV)}&
\colhead{$EW$\tablenotemark{c}  (eV)}&
\colhead{$\chi_{red}^2$ ($\chi^2$/$N_{dof}$)}}
\startdata
1 & 24/08/2002\tablenotemark{*} & ~SIMS & 2.37$\pm$0.02 & 0.22$\pm$0.01 & 1.29$\pm$0.07 & 233$\pm$40 & 1.19 (2318/1941) \\   
2& 29/09/2002\tablenotemark{**}  & ~HSS & 2.54$\pm$0.01 & 0.30$\pm$0.01 & 1.24$\pm$0.02 & 684$\pm$20 & 1.27 (2469/1941) \\   
3 & 08/03/2003\tablenotemark{***} & ~HIMS & 2.17$\pm$0.02 & 0.38$\pm$0.01 & 0.69$\pm$0.11 & 167$\pm$55 & 1.03 (1997/1945) \\   
\enddata
\tablenotetext{a}{BH states: SIMS - soft intermediate state, HIMS - hard intermediate state, HSS - high soft state \citep[see ][ for details]{hb05}.} 
\tablenotetext{b}{$f_C=A/(1+A).$} 
\tablenotetext{c}{The centroid energy of the Gaussian was  fixed at 6.4 keV.}
\tablenotetext{*}{XMM/{\it RXTE} Obs ID: 0093562701/70130-01-01-00}
\tablenotetext{**}{XMM/{\it RXTE} Obs ID: 0156760101/70130-01-02-00}
\tablenotetext{***}{XMM/{\it RXTE} Obs ID: 0148220201/50117-01-03-00 }
\label{gx339_tab}
\end{deluxetable}
\newpage

\newpage
\begin{deluxetable}{lllll}
\tablewidth{0pt}
\tabletypesize{\footnotesize}
\tablecaption{Best-fit Parameters for the  XMM/{\it RXTE} Spectrum of BH GX 339-4 on September 29, 2002 (Observation 2)}
\tablehead{
\colhead{Component} &
\colhead{Parameter} &
\colhead{LAOR\tablenotemark{a}} &
\colhead{DISKLINE} &
\colhead{WINDLINE}}
\startdata
WABS&$N_H$ (cm$^2$) & 0.354$\pm$0.005 & 0.356$\pm$0.005 & 0.365$\pm$0.006 \\
BBODY&$T_{bb}$ (keV) & 0.332$\pm$0.003 & 0.331$\pm$0.003 & 0.319$\pm$0.004 \\
BMC&$kT$ (keV) & 0.679$\pm$0.003 & 0.678$\pm$0.002 & 0.662$\pm$0.004 \\
&$\alpha$  & 1.55$\pm$0.01 & 1.60$\pm$0.01 & 1.44$\pm$0.02 \\
&$\log (A)$  & -0.74$\pm$0.01 & -0.75$\pm$0.01 & -0.91$\pm$0.01 \\
K$_\alpha$ line&$E_L$ (keV) & 6.4\tablenotemark{b} & 6.4\tablenotemark{b} & 6.60$\pm$0.15 \\
&$\beta_L$ & 1.62$\pm$0.04 & 2.40$\pm$0.05 & - \\
&$R_{in}$ ($R_G$)  & 1.23\tablenotemark{b} & 6.0\tablenotemark{b} & - \\
&$R_{out}$ ($R_G$)  & 400\tablenotemark{f} & 1000\tablenotemark{f} & - \\
&$i_L$ (deg)  & 80\tablenotemark{c} & 80\tablenotemark{c} & - \\
&$\tau_W$ & - & - & 4.9$\pm$0.2 \\
&$\beta_W=V_W/c$  & - & - & (2.47$\pm$0.02)$\times 10^{-2}$ \\
&$kT_W$ (keV) & - & - & 1.06$\pm$0.02 \\
\hline
&$\chi^2_{red}$ ($\chi^2$/$N_{dof}$) & 1.33 (2573/1937) & 1.34 (2596/1937) & 1.24  (2401/1937)\\
\enddata
\tablenotetext{a}{Model is specified by the name at the top of a column} 
\tablenotetext{b}{Parameter value pegged into the lower limit set by the model} 
\tablenotetext{c}{Parameter value pegged into the upper limit set by the model} 
\tablenotetext{f}{Parameter fixed}
\label{gx339_sept292002_tab}
\end{deluxetable}
\clearpage
\newpage



\newpage

\newpage
\begin{figure}[ptbptbptb]
\includegraphics[scale=0.7, angle=-90]{f3a.eps}

\includegraphics[scale=0.705, angle=-90]{f3b.eps}
\caption{The XMM-{\it Newton} observation of  GK Per  (rev. 412) 
The data (error bars), model (solid line).  Here we have used the best-fit  model wabs$\ast$(COMPTT+BBODY+LAOR)$\ast$EDGE (see Table 1).}
\label{GKPER_fit_laor}
\end{figure}

\begin{figure}[ptbptbptb]
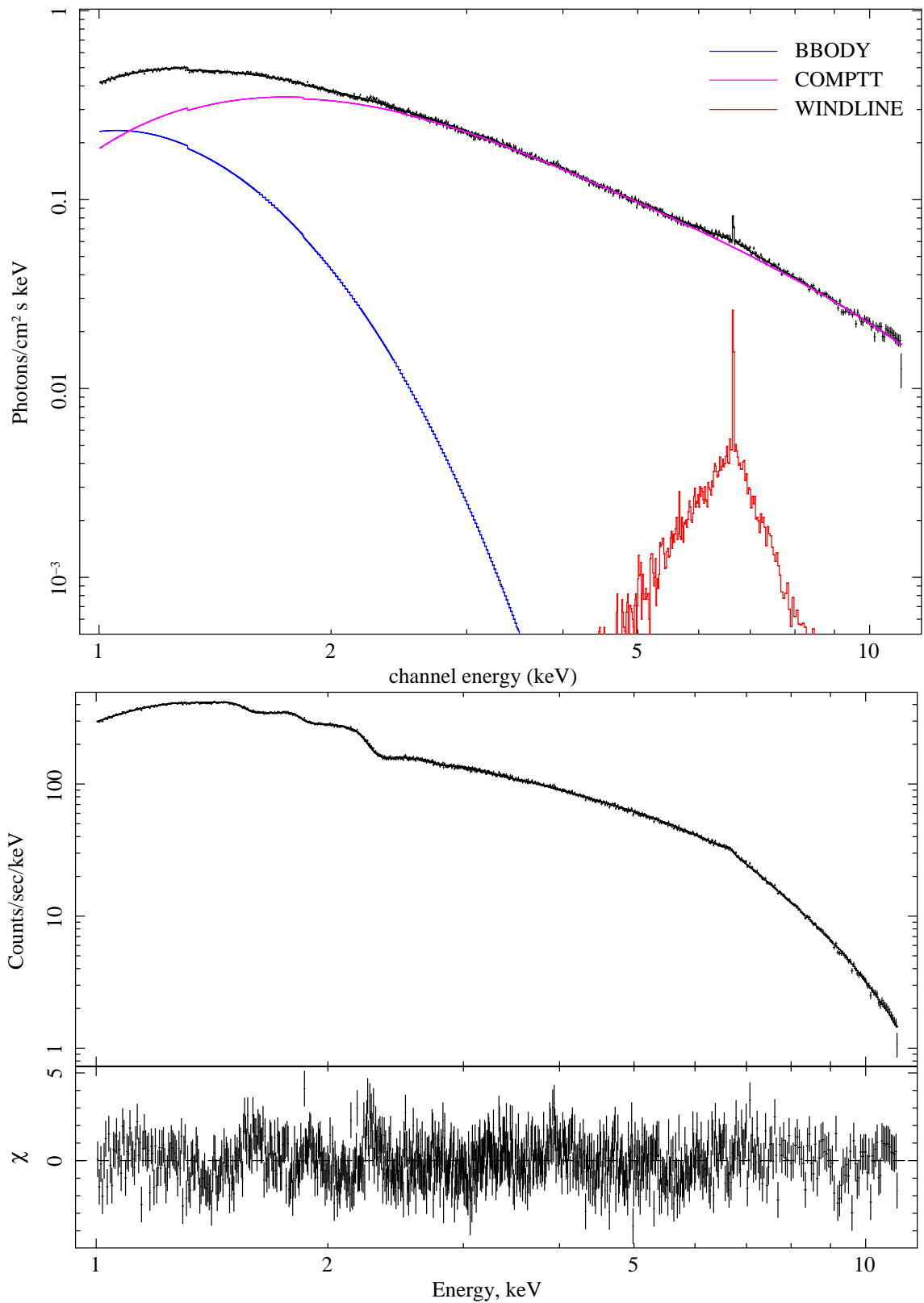

\includegraphics[scale=0.65, angle=-90]{f4a.eps}
\includegraphics[scale=0.65, angle=-90]{f4b.eps}
\caption{The XMM-{\it Newton} observation of  Ser X-1 (XMM rev. 785) 
 The upper panel shows the data (error
bars), model (solid line).   Here we have used the best-fit XSPEC model 
WABS$\ast$(COMPTT+BBODY+WINDLINE) (see Table \ref{serpx1_tab}).
The lower panel shows the count spectrum along the residual of the data vs model in units of   $\chi=\rm{(data-model)}/\sigma$.
}
\label{ser_fit}
\end{figure}

\begin{figure}[ptbptbptb]
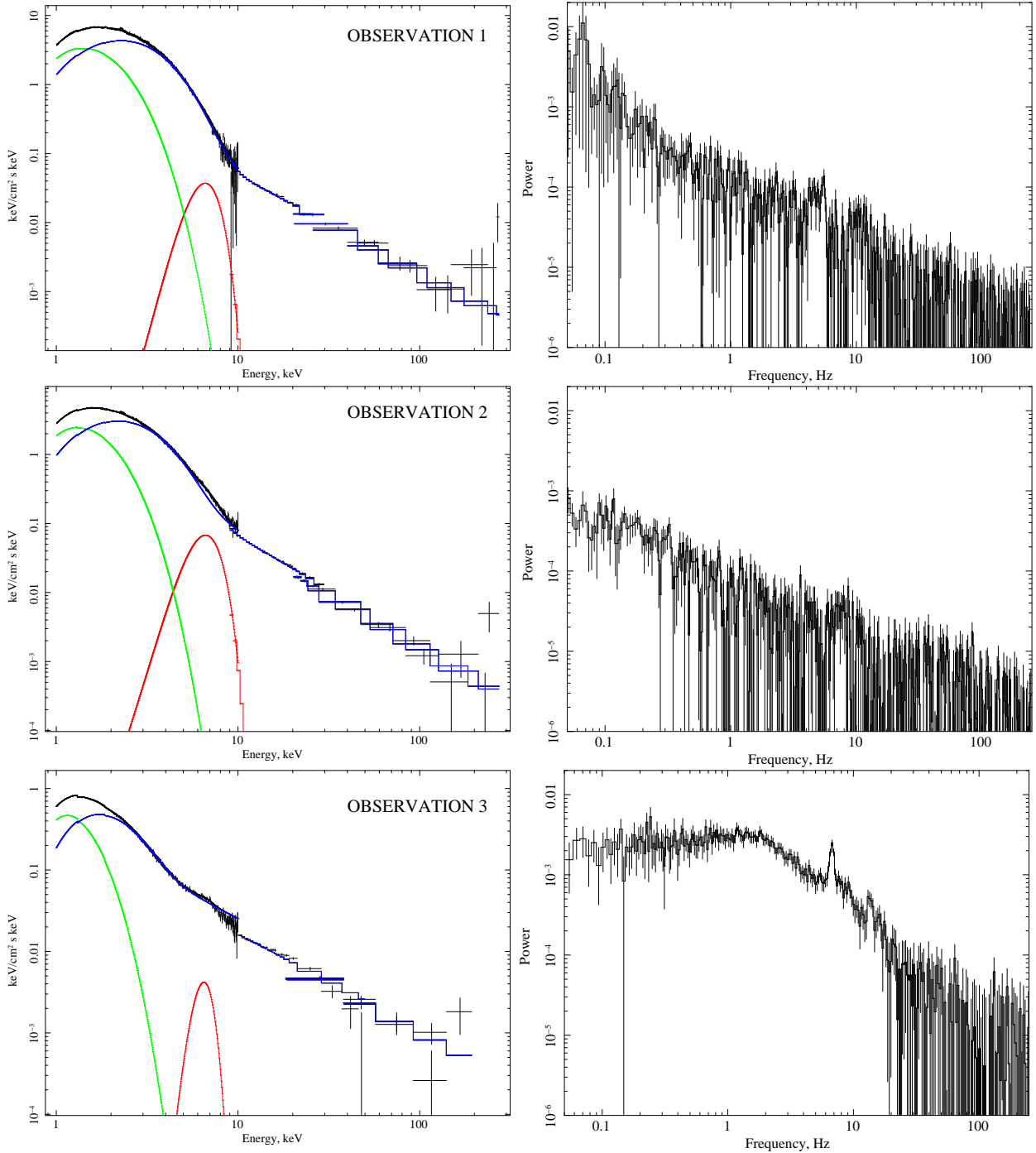

\includegraphics[scale=0.35, angle=-90]{f5a.eps}
\includegraphics[scale=0.35, angle=-90]{f5b.eps}
\includegraphics[scale=0.35, angle=-90]{f5c.eps}
\includegraphics[scale=0.35, angle=-90]{f5d.eps}
\includegraphics[scale=0.35, angle=-90]{f5e.eps}
\includegraphics[scale=0.35, angle=-90]{f5f.eps}
\caption{X-ray spectra (left panels) and PDS (right panels) for the XMM-{\it Newton}/{\it RXTE} observations.
Note that PDSs  shows the variability features (white-red noise, QPO and break frequencies) when the line is weak (see bottom panels). On the other hand PDSs are featureless when the line is strong (see top and middle panels).}
\label{gx339_ga_eufs}
\end{figure}

\begin{figure}[ptbptbptb]
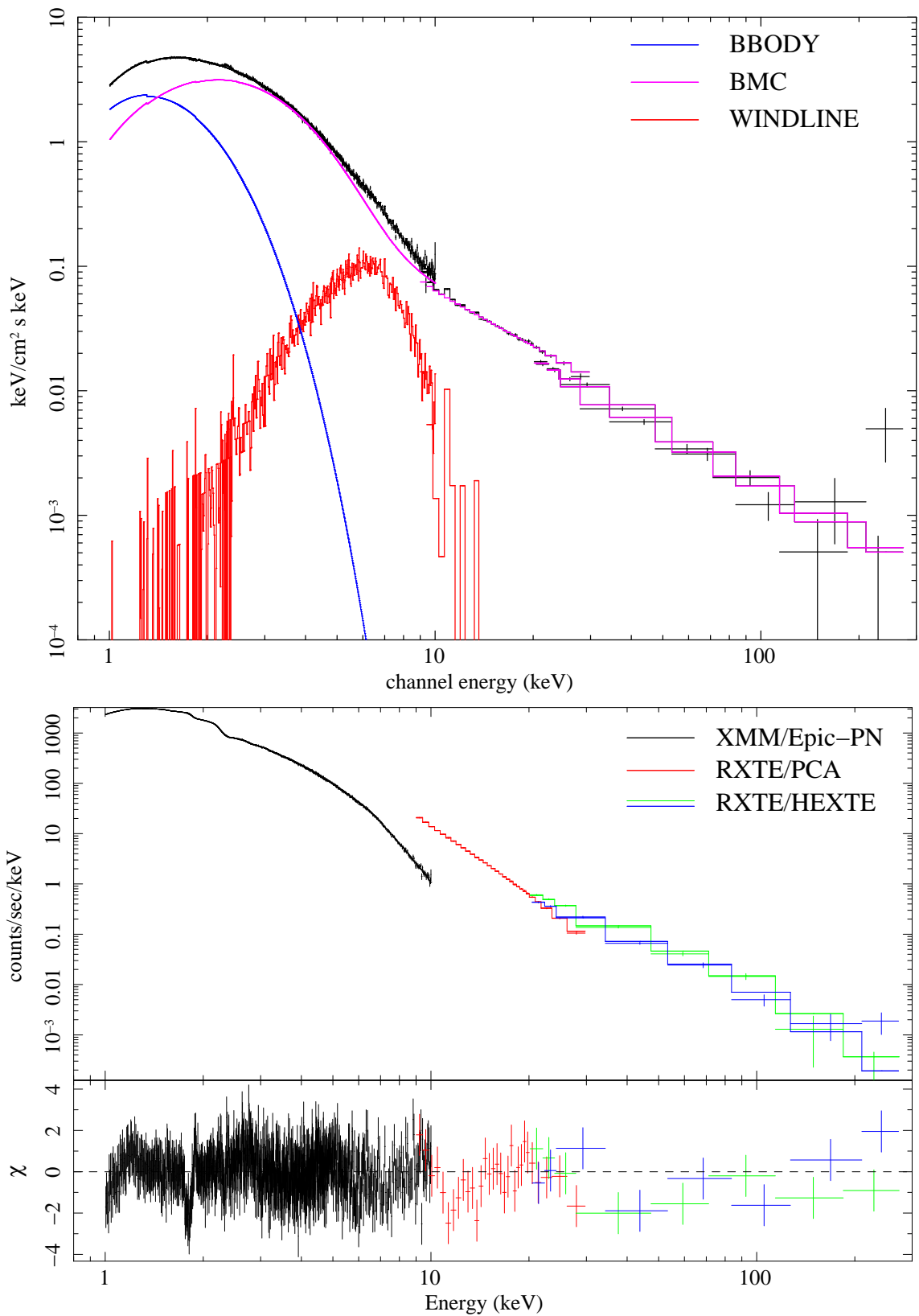

\includegraphics[scale=0.65, angle=-90]{f6a.eps}
\includegraphics[scale=0.65, angle=-90]{f6b.eps}
\caption{The XMM-{\it Newton} and {\it RXTE} joint 
 observation of  GX 339-4 on September 29, 2002 (Observation 2).
Here we have used the best-fit XSPEC model wabs$\ast$(BMC+BB+GAUSSIAN+WINDLINE) (see Table \ref{gx339_sept292002_tab}).
In the top panel we show unfolded specrtum where different colors indicate different model components.
The bottom panel shows count spectrum with data from different instruments shown by different 
colors. 
}
\label{gx339_514_fit}
\end{figure}

\begin{figure}[ptbptbptb]
\includegraphics[width=6.2in,height=6.2in, angle=-90]{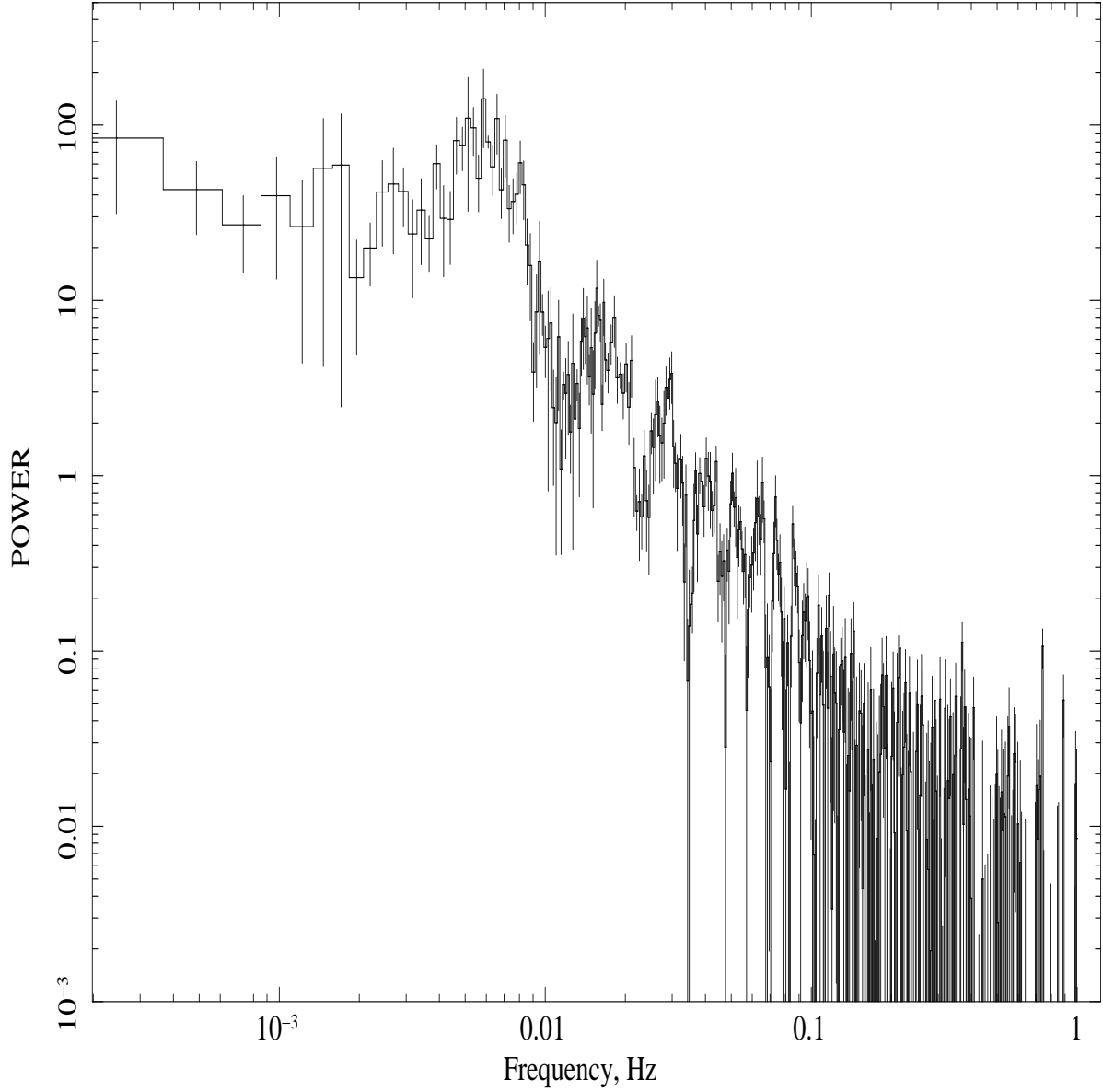}
\caption{XMM PDS related to GK Per photon spectrum (see Fig. \ref{GKPER_fit}).
One clearly see that there is no variability for frequencies higher than 0.03 Hz.}
\label{pds_gkper}
\end{figure}





\begin{figure}[ptbptbptb]
\includegraphics[width=8.2in,height=6.2in, angle=-90]{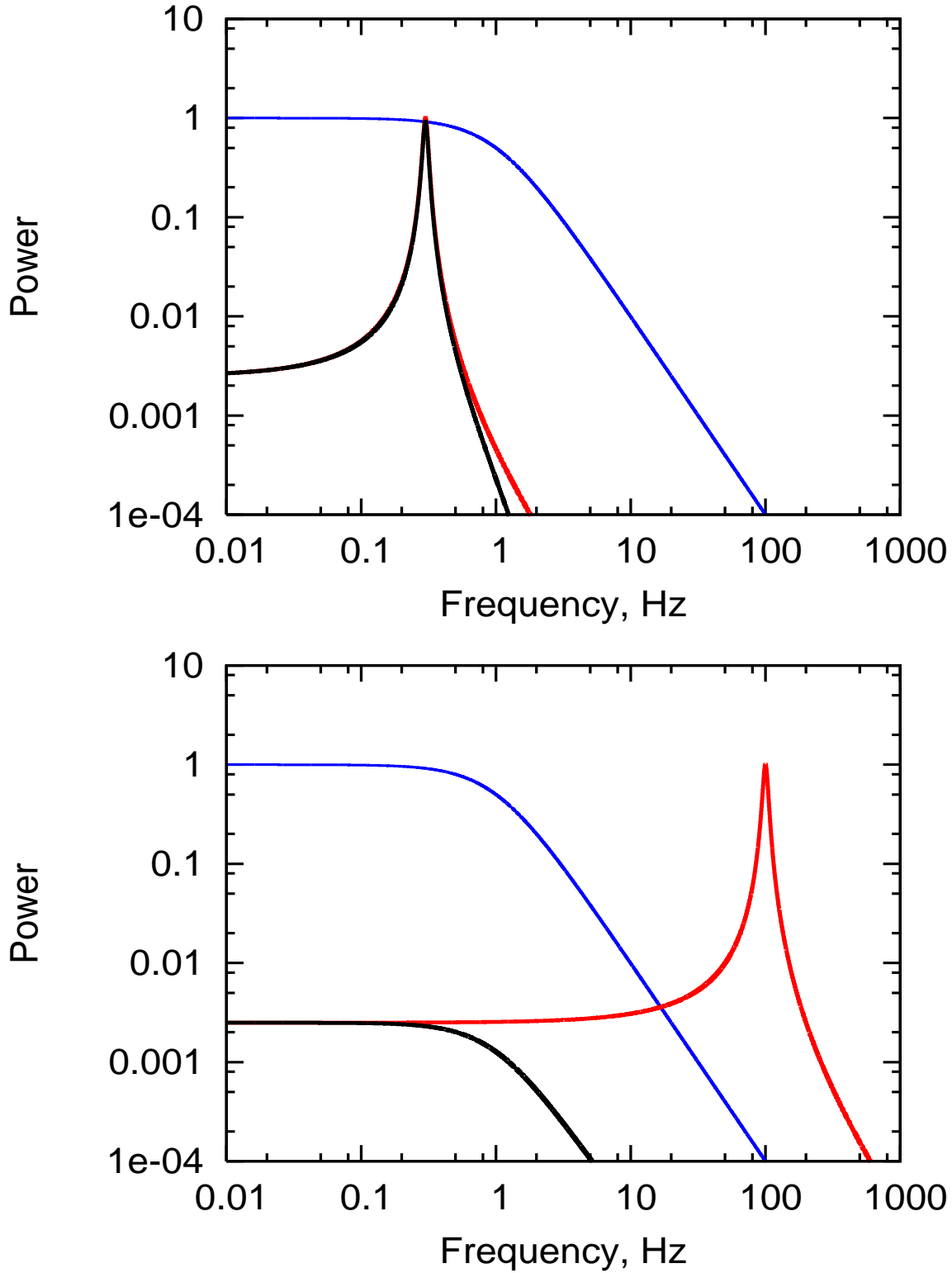}
\caption{ Photon scattering effect in the wind cloud. Blue line stands for the PDS related to the wind response (exponential shot) due to photon scattering. Red line describes the QPO PDS profile.
Black line is the resulting PDS as a product of the wind PDS and the QPO Lorentzian.
{\it Upper panel}: Low frequency is unaffected by the photon scattering in wind.
{\it Lower panel}: the QPO features are completely washed in the wind if a characteristic QPO frequency is much higher that of the wind.
 }
\label{photon_scat_pds}
\end{figure}
\begin{figure}[ptbptbptb]
\includegraphics[width=6.2in,height=6.2in, angle=0]{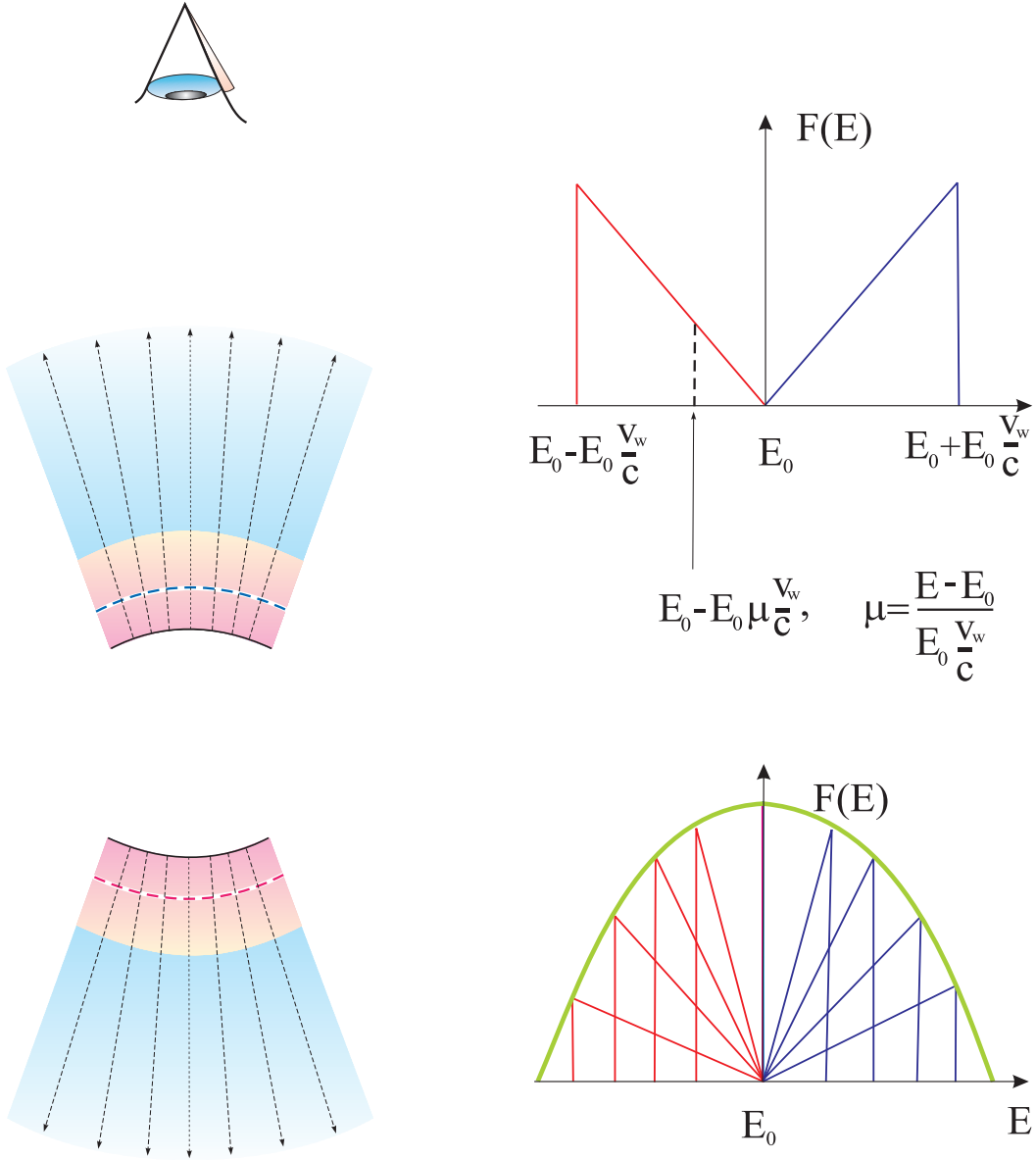}
\caption{Spherical wind geometry and wind line profiles of direct component. 
 }
\label{wind_geom}
\end{figure}

\end{document}